\documentclass[12pt]{article} 

\usepackage{amsmath,amssymb,amsfonts}
\usepackage{psfrag}
\usepackage{color}
\definecolor{darkblue}{rgb}{0.1,0.1,.7}
\usepackage[square, comma, compress,numbers]{natbib}
\usepackage[]{amsmath}
\usepackage[]{graphicx}
\usepackage[]{latexsym}
\usepackage{geometry}
\usepackage{amscd}
\usepackage[all,cmtip]{xy}
\usepackage{mathrsfs}
\usepackage[margin=10pt,font=small,labelfont=bf]{caption}
\usepackage[colorlinks, linkcolor=darkblue, citecolor=darkblue, urlcolor=darkblue, linktocpage]{hyperref} 
\usepackage{simplewick}
\usepackage{bbold}
\usepackage{changepage}
\usepackage{youngtab}
\usepackage{float}
\numberwithin{equation}{section}

\newcommand{\be}{\begin{eqnarray}}
\newcommand{\ee}{\end{eqnarray}}
\newcommand{\bea}{\begin{eqnarray}}
\newcommand{\eea}{\end{eqnarray}}

\newcommand   \Tr    {\mathrm{Tr}}

\newcommand   \f  {\phi}

\def\beq{\begin{equation}} 
\def\eeq{\end{equation}} 
\def\<{\langle}
\def\>{\rangle}

\def\nn{\nonumber} 
 
\def\cO {{\cal O}} 
\def\cM {{\cal M}}

\def\cN {{\cal N}}

\begin{document}

\vspace*{-.6in} \thispagestyle{empty}
\begin{flushright}
\end{flushright}
\vspace{.2in} {\Large
\begin{center}
{\bf Non-Abelian Binding Energies from the \\Lightcone Bootstrap\\\vspace{.1in}}
\end{center}
}
\vspace{.2in}

\begin{center}
{\bf 
Daliang Li$^{a,b}$, David Meltzer$^a$, David Poland$^{a,c}$} 
\\
\vspace{.2in} 
$^a$ {\it  Department of Physics, Yale University, New Haven, CT 06511}\\
$^b$ {\it  Department of Physics and Astronomy, Johns Hopkins University, Baltimore, MD 21218}\\
$^c$ {\it School of Natural Sciences, Institute for Advanced Study, Princeton, NJ 08540}
\end{center}

\vspace{.2in}

\begin{abstract}
We analytically study the lightcone limit of the conformal bootstrap for 4-point functions containing scalars charged under global symmetries. 
We show the existence of large spin double-twist operators in various representations of the global symmetry group. 
We then compute their anomalous dimensions in terms of the central charge $C_T$, current central charge $C_J$, and the OPE coefficients of low dimension scalars. 
In AdS, these results correspond to the binding energy of two-particle states arising from the exchange of gravitons, gauge bosons, and light scalar fields. 
Using unitarity and crossing symmetry, we show that gravity is universal and attractive among different types of two-particle states, while the gauge binding energy can have either sign as determined by the representation of the two-particle state, with universal ratios fixed by the symmetry group. We apply our results to 4D $\mathcal{N}=1$ SQCD and the 3D $O(N)$ vector models. We also show that in a unitary CFT, if the current central charge $C_J$ stays finite when the global symmetry group becomes infinitely large, such as the $N\rightarrow\infty$ limit of the $O(N)$ vector model, then the theory must contain an infinite number of higher spin currents.
\end{abstract}

\newpage

\tableofcontents

\newpage

\section{Introduction}
\label{sec:intro}

Recently there has been a resurgence of interest in the conformal bootstrap~\cite{Polyakov:1974gs,Ferrara:1973yt,Mack:1975jr} approach to studying CFTs in $D>2$, including rigorous numerical bounds on scaling dimensions and OPE coefficients~\cite{Rattazzi:2008pe,Rychkov:2009ij,Caracciolo:2009bx,Poland:2010wg,Rattazzi:2010gj,Rattazzi:2010yc,Vichi:2011ux,Poland:2011ey,Rychkov:2011et,ElShowk:2012ht,Liendo:2012hy,Beem:2013qxa,Kos:2013tga,El-Showk:2013nia,Alday:2013opa,Gaiotto:2013nva,Bashkirov:2013vya,Berkooz:2014yda,Nakayama:2014lva,Nakayama:2014yia,Alday:2014qfa,Kos:2014bka,Chester:2014fya,Caracciolo:2014cxa,Nakayama:2014sba,Golden:2014oqa,Paulos:2014vya,Bae:2014hia,Beem:2014zpa,Chester:2014gqa,Simmons-Duffin:2015qma,Bobev:2015vsa,Bobev:2015jxa,Kos:2015mba,Iliesiu:2015qra,Beem:2015aoa,Chester:2015qca,Poland:2015mta,Lemos:2015awa}, methods for constructing approximate solutions to crossing symmetry~\cite{ElShowk:2012hu,Gliozzi:2013ysa,Gliozzi:2014jsa,Gliozzi:2015qsa} leading to high-precision determinations of the operator spectrum in the 3D Ising~\cite{ElShowk:2012ht,El-Showk:2014dwa,Kos:2014bka,Simmons-Duffin:2015qma} and $O(N)$ vector models~\cite{Kos:2013tga,Kos:2015mba}, and new insights into supersymmetric CFTs, including 3D $\cN=1$~\cite{Bashkirov:2013vya,Iliesiu:2015qra}, $\cN=2$~\cite{Bobev:2015vsa,Bobev:2015jxa,Chester:2015qca}, and $\cN=8$~\cite{Chester:2014fya,Chester:2014mea} theories, 4D $\cN=1$~\cite{Poland:2010wg,Vichi:2011ux,Poland:2011ey,Berkooz:2014yda,Poland:2015mta}, $\cN=2$~\cite{Beem:2013sza,Beem:2014rza,Beem:2014zpa,Lemos:2015awa}, and $\cN=4$~\cite{Beem:2013qxa,Alday:2013opa,Alday:2014qfa,Alday:2014tsa} theories, and the mysterious 6D SCFTs with $(2,0)$ supersymmetry~\cite{Beem:2014kka,Beem:2015aoa}. 

In addition to these studies, very general analytical constraints have been obtained by considering the bootstrap equations in the lightcone limit. In particular, it was argued in~\cite{Fitzpatrick:2012yx,Komargodski:2012ek} and extended in~\cite{Alday:2013cwa,Costa:2013zra,Fitzpatrick:2014vua,Vos:2014pqa,Kaviraj:2015cxa,Alday:2015eya,Kaviraj:2015xsa,Fitzpatrick:2015qma,Alday:2015ota} that the bootstrap conditions imply that every CFT containing scalars $\phi_i$ of dimension $\Delta_{i}$ must contain towers of ``double-twist" operators $\cO_{\tau,\ell}$ in the $\phi_i \times \phi_j$ OPE with twist $\tau \rightarrow \Delta_{i} + \Delta_{j} + 2n + \frac{\gamma_n}{\ell^{\tau_m}}$ for integer $n$ as their spin $\ell \rightarrow \infty$. Here $\tau_m$ is the minimal twist in the $\phi \times \phi$ OPE and the coefficients $\gamma_n$ can be calculated in terms of the OPE coefficient of the leading twist operator. This general structure was anticipated in the earlier work of~\cite{Alday:2007mf}. Other nontrivial constraints on CFT correlators in the Lorentzian regime have been recently studied in~\cite{Camanho:2014apa, Maldacena:2015waa, Hartman:2015lfa,Maldacena:2015iua}.

In many CFTs this leading twist operator is the stress-energy tensor, in which case these corrections have a simple interpretation in the AdS/CFT correspondence -- they are simply the gravitational binding energies of two-particle states of large spin, which are expected to be negative $\gamma_n < 0$ due to the attractive nature of gravitational interactions. If the $\phi_i$ are charged under a global symmetry, then currents can also appear in the OPE, corresponding to gauge interactions in the bulk. The binding energy of large spin two-particle states from gauge interactions can be computed via crossing symmetry. For example, in the case of a complex scalar $\phi$ charged under an Abelian symmetry, one can show that the the double-twist states in the $\phi \times \phi$ OPE receive positive corrections (corresponding to repulsive like-charge interactions in AdS) while the double-twist states in the $\phi^{\dagger} \times \phi$ OPE receive negative corrections of the same size (corresponding to attractive opposite-charge interactions in AdS)~\cite{Komargodski:2012ek}. 

This analysis can be extended to non-Abelian symmetries. For example, two scalars in the fundamental representation of $SU(N)$ can combine into families of two-particle states transforming under the symmetric or anti-symmetric representations with binding energies denoted by $\gamma^S$ and $\gamma^A$. 
We will compute them as a function of the current central charge $C_J$ of the CFT. 
We will see that their sign is determined only by the representation of the two-particle states, and that $\gamma^S/\gamma^A$ is a function of $N$ that is independent of the dynamic details of the theory. 
Using crossing symmetry, it will be obvious that these features holds for any symmetry group. We will also compute the binding energies when there are $O(N)$ and $SU(N)$ global symmetries with fundamentals, adjoints, and symmetric tensors.

Perhaps less intuitive are the consequences of charged operator exchange -- e.g., in the case of a $U(1)$ symmetry we will see that exchanging a charged scalar induces corrections to the anomalous dimensions that are negative for even spins but positive for odd spins. More generally, for non-Abelian global symmetries, exchanging charged operators induces an intricate set of representation and spin-dependent corrections to the dimensions of double-twist states. We will work out these corrections for $O(N)$ and $SU(N)$ global symmetries with fundamentals, adjoints, and symmetric tensors. We apply these results to 4D $\cN=1$ SQCD and the 3D $O(N)$ vector models. 

All the results outlined above do not rely on any type of large $N$ limit. However, we will show in section \ref{subsec:largeN} that crossing symmetry and unitarity implies a sufficient condition for the existence of higher spin symmetries in the large $N$ limit of CFTs with $O(N)$ or $SU(N)$ global symmetries.
In particular, we predict that if a unitary CFT has $C_J\sim\mathcal{O}(1)$ as $N \rightarrow \infty$, then in this limit it must have higher spin conserved currents in order to solve the crossing equations, while if $C_J \geq \mathcal{O}(N)$ they are not required. In the former case, we also argue that the theory cannot contain scalar operators with dimension $\Delta < d-2$ whose coefficients remain $\mathcal{O}(1)$ at large $N$. Examples of theories in both these classes can be found. 

While the results of this analysis may have applications to many other theories, we also view this as an important precursor to the more sophisticated analysis of correlation functions containing global symmetry currents and stress-energy tensors, which we pursue in a separate publication~\cite{CurrentPaper:2015}.

This paper is organized as follows. In Section~\ref{sec:main} we briefly review the argument that allows us to compute the dimensions of double-twist operators and apply this reasoning to CFTs with $U(1)$, $O(N)$, and $SU(N)$ global symmetries. In Section~\ref{sec:applications} we discuss some applications of our results and in Section~\ref{sec:conclusions} we conclude. Appendix~\ref{app:scalarOPE} briefly reviews the scalar-scalar OPE. Appendix~\ref{app:structs} contains further details about the tensor structures and crossing relations that we suppress in the main text. Appendix~\ref{app:LargeNArgument} provides some technical details for the analysis at large $N$.

\section{From Crossing Equations to Binding Energies}
\label{sec:main}

\subsection{Real Scalars}

We will start by reviewing the basic results of~\cite{Fitzpatrick:2012yx,Komargodski:2012ek} and establishing some notation. Let us consider a CFT 4-point function containing two real scalars $\phi_{1}$ and $\phi_{2}$ of the form $\langle\phi_{1}(x_1)\phi_{1}(x_2)\phi_{2}(x_3)\phi_{2}(x_4)\rangle$. Expanding this 4-point function in conformal blocks and equating the $s$-channel and $t$-channel expansions gives a crossing relation of the form
\begin{equation}\label{eq:Crossingforf1f2}
\sum_{\cO \in \phi_{1,2} \times \phi_{1,2}} P^{11,22}_{\cO}  g^{11,22}_{\tau,\ell}(u,v) = u^{\Delta_2}v^{-\frac{1}{2}(\Delta_1+\Delta_2)} \sum_{\cO \in \phi_1 \times \phi_2} P^{12,21}_{\cO} g^{12,21}_{\tau,\ell}(v,u),
\end{equation}
where the coefficients are related to the OPE coefficients as $P^{ij,kl}_{\cO} \equiv \left(\frac{-1}{2}\right)^{\ell} \lambda_{\phi_i \phi_j \cO} \lambda_{\phi_k \phi_l \cO}$, we label the conformal blocks $g^{ij,kl}_{\tau,\ell}(u,v)$ by the twist $\tau$ and spin $\ell$ of the exchanged operator, and we work in a normalization such that $g_{\tau,\ell}(u,v) \rightarrow u^{\tau/2} (1-v)^{\ell}$ when we take $u \rightarrow 0$ and then $v \rightarrow 1$. The twist of an operator is defined as $\tau=\Delta-\ell$, where $\Delta$ is its conformal dimension.

To satisfy the crossing equations, the $\phi_1 \times \phi_2$ OPE should contain a tower of double-twist operators that are schematically of the form $\cO_{n,\ell} \sim \phi_{1}\partial^{\ell}\partial^{2n}\phi_{2}$ and have twist approaching $\tau \rightarrow \Delta_1 + \Delta_2+2n$ as $\ell \rightarrow \infty$. One can see this rigorously by considering (\ref{eq:Crossingforf1f2}) in the eikonal limit $u \ll v \ll 1$. As in Section 2.3 of~\cite{Fitzpatrick:2012yx}, we will make the assumption that this tower is isolated in the sense that there is a single operator at each value of $n$ and $\ell$ that gives the dominant contribution to the large-$\ell$ sum in the 4-point function. In this work we will focus our attention on the $n=0$ tower with lowest twist $\cO_{\ell} \equiv \cO_{0,\ell}$, though all of the results can be straightforwardly extended to larger values of $n$ following~\cite{Kaviraj:2015cxa,Kaviraj:2015xsa}.

Under this assumption, the OPE coefficients and anomalous dimensions $\gamma_{\ell}$ of these operators can then be calculated using the conformal bootstrap equations by matching the infinite sum over spins on the RHS to the singularities of the minimal-twist contributions that are shared between the $\phi_1 \times \phi_1$ and $\phi_2 \times \phi_2$ OPEs on the LHS, corresponding to the approximate relation
\begin{equation}
1 + P^{11,22}_{\cO_m}  g^{11,22}_{\tau_m,\ell_m}(u,v) + \ldots \approx u^{\Delta_2}v^{-\frac{1}{2}(\Delta_1+\Delta_2)}  \sum_{\ell}  P_{\cO_{\ell}}^{12,21} g^{12,21}_{\Delta_1 + \Delta_2 + \gamma_{\ell},\ell}(v,u).
\end{equation}
 The leading contributions on the LHS arise from the identity operator, and the next contribution could either be a low-dimension scalar or the stress-energy tensor with twist $\tau_{m}=d-2$, where $d$ is the spacetime dimension.

To be more specific, as described in~\cite{Fitzpatrick:2012yx,Komargodski:2012ek}, matching the identity operator contribution on the LHS to the infinite sum over spins on the RHS in the lightcone limit $u \ll v \ll 1$ fixes the leading behavior of the OPE coefficient to be
\be
P^{12,21}_{\cO_{\ell}}=\frac{2^{2-\Delta_{1}-\Delta_{2}}\sqrt{\pi}}{\Gamma(\Delta_{1})\Gamma(\Delta_{2})} \frac{\ell^{\Delta_1+\Delta_2-\frac32}}{2^{2\ell}} \equiv P_{\Delta_1,\Delta_2}(\ell).
\ee

Next, by matching the $\log(v)$ singularity contained in the conformal block of the minimal twist (non-identity) operator $\cO_m$ on the LHS to the $\log(v)$ obtained by expanding the anomalous dimensions of the $\cO_{\ell}$ operators on the RHS gives
\begin{equation}
\gamma_{\ell}=-P_{\cO_{m}}^{11,22}\frac{\xi_{\Delta_1,\Delta_2}^{\cO_{m}}}{\ell^{\tau_m}},
\end{equation}
where the coefficient
\begin{equation}
\xi_{\Delta_1,\Delta_2}^{\cO_{m}}\equiv\frac{2\Gamma(\Delta_{1})\Gamma(\Delta_{2})\Gamma(\tau_m+2\ell_{m})}{\Gamma(\Delta_{1}-\frac{\tau_m}{2})\Gamma(\Delta_{2}-\frac{\tau_m}{2})\Gamma(\frac{\tau_m}{2}+\ell_{m})^{2}}
\end{equation}
is a positive quantity. In general the correction to the anomalous dimension could have either sign due to the product of different OPE coefficients, but in the case of stress-tensor exchange the coefficients are fixed by the Ward identity to have the same sign, leading to a negative-definite anomalous dimension. Note that the unitarity bound is $\Delta_{i} \geq \frac{d}{2}-1$ for scalars and $\tau\geq d-2$ for operators with spin. Therefore, when the exchanged operator is conserved, the $\Gamma$ functions in the denominator force the anomalous dimensions to vanish if either $\phi_1$ or $\phi_2$ is a free field.

In the next sections we will generalize this matching to cases where the scalars are charged under a global symmetry.  The general form of the crossing relations in this situation appeared in~\cite{Rattazzi:2010yc} and some aspects of this situation in the context of the lightcone bootstrap were discussed in~\cite{Komargodski:2012ek}. Here we will give a more detailed analysis, taking care to disentangle the double-twist operators in different global symmetry representations.

\subsection{Complex Scalars}

To begin, we start with a complex scalar $\phi$ charged under a $U(1)$ global symmetry and consider 4-point functions of the form $\langle\phi(x_{1})\phi^{\dagger}(x_{2})\phi(x_{3})\phi^{\dagger}(x_{4})\rangle$. Relating the $(12)-(34)$ channel to the $(14)-(32)$ channel in the lightcone limit gives the relation
\begin{equation}\label{eq:complex1}
1+ P_{\epsilon} g_{\tau_{\epsilon},0}(u,v) + P_J g_{d-2,1}(u,v)+P_T g_{d-2,2}(u,v)\approx\left(\frac{u}{v}\right)^{\Delta_{\phi}}{\displaystyle \sum_{\ell}} P_{\cO_{\ell}} g_{2\Delta_{\phi}+ \gamma_{\ell},\ell}(v,u),
\end{equation}
where we explicitly show the contributions of the leading scalar $\epsilon$, the $U(1)$ global symmetry current $J_\mu$, and the stress-energy tensor $T_{\mu\nu}$ on the LHS. The coefficients are written in terms of the positive quantities $P_{\cO} \equiv \frac{1}{2^{\ell}} |\lambda_{\phi \phi^{\dagger} \cO}|^2$ and the RHS runs over double-twist operators of the form $\cO_{\ell} \sim \phi\partial^{\ell} \phi^{\dagger}$. 

Alternatively, we can relate the $(12)-(43)$ channel to the $(13)-(42)$ channel, giving the sum rule
\begin{equation}\label{eq:complex2}
1+P_{\epsilon} g_{\tau_{\epsilon},0}(u,v)-P_J g_{d-2,1}(u,v)+P_T g_{d-2,2}(u,v)\approx\left(\frac{u}{v}\right)^{\Delta_{\phi}}{\displaystyle \sum_{\ell^+}}P^+_{\cO_{\ell}^+} g_{2\Delta_{\phi} + \gamma^+_{\ell},\ell}(v,u),
\end{equation}
where we have implicitly relabeled the coordinates so the crossing symmetry equations take the same form as before. In this case the current has an opposite sign on the LHS and the RHS runs over charged double-twist operators of the form $\cO^+_{\ell} \sim \phi \partial^{\ell} \phi$ with coefficients $P^+_{\cO_{\ell}^+} \equiv \frac{1}{2^{\ell}} |\lambda_{\phi\phi\cO^+_{\ell}}|^{2}$. Here the notation $\ell^+$ means that the sum only runs over even spins.

Finally, if there is a low-twist charged scalar $c$ exchanged in the $\phi \times \phi$ OPE then by switching the role of $u$ and $v$ we also have the condition
\be\label{eq:complex3}
P^+_{c} g_{\tau_c,0}(u,v) \approx \left(\frac{u}{v}\right)^{\Delta_{\phi}} \sum_{\ell} P_{\cO_{\ell}} (-1)^{\ell} g_{2\Delta_{\phi} + \gamma_{\ell},\ell}(v,u).
\ee

Matching the identity in (\ref{eq:complex1}) and (\ref{eq:complex2}) yields the mean-field theory behavior
\begin{equation}
P_{\cO_{\ell}}=\frac12P^+_{\cO_{\ell}^+} = P_{\Delta_{\phi},\Delta_{\phi}}(\ell),
\end{equation}
while matching terms of order $u^{\frac{d-2}{2}}\log(v)$ on both sides gives the shifts in the anomalous dimensions
\begin{equation}\label{eq:GammaTComplexScalar}
\delta_T \gamma_{\ell}= \delta_T \gamma_{\ell}^+=-\frac{d^2 \Delta_{\phi}^2}{4(d-1)^2 C_{T}S_{d}^{2}} \frac{\xi_{\Delta_{\phi},\Delta_{\phi}}^T}{\ell^{d-2}}
\end{equation}
arising from stress-tensor exchange, and the shifts
\begin{equation}\label{eq:GammaJComplexScalar}
\delta_J \gamma_{\ell} = - \delta_J \gamma_{\ell}^+=-\frac{1}{2 C_{J}S_{d}^{2}} \frac{\xi_{\Delta_{\phi},\Delta_{\phi}}^{J}}{\ell^{d-2}}
\end{equation}
arising from current exchange. We have inserted the value of $P_T$ and $P_J$ as determined by the Ward identity where $S_{d}=\frac{2\pi^{\frac{d}{2}}}{\Gamma(\frac{d}{2})}$ is the area of the d-1 dimensional sphere. Our normalization of the conserved currents and stress energy tensor differ by a factor of $S_{d}^{2}$ in comparison to previous work on the conformal bootstrap (see Appendix~\ref{app:scalarOPE} for our conventions). The corrections from the stress-tensor are universal and negative while the corrections due to current exchange to the two double-twist states have opposite signs but the same magnitude. 

In the bulk, the anomalous dimensions (\ref{eq:GammaTComplexScalar}) and (\ref{eq:GammaJComplexScalar}) correspond to the binding energies between a pair of well separated particles arising from gravitational and gauge interactions. If the weak gravity conjecture holds, then there should exist a particle in the bulk for which the gravitational attraction is dominated by the $U(1)$ gauge repulsion, or $\delta_T \gamma_{\ell}^+ +\delta_J \gamma_{\ell}^+> 0$. Note that we assumed the operator $\phi$ to have unit charge. If we consider instead an operator $\phi_q$ with carrying charge $q$, then this condition holds if 
\be
\frac{\Delta_{\phi_q}^2}{q^2} < \frac{(d-1)^2}{2d(d+1)} .
\ee
At $d=4$, this matches with the kinematic version of the weak gravity conjecture found in~\cite{Nakayama:2015hga}.

We also find the contribution to the anomalous dimensions from the exchange of a light scalar by matching terms of order $u^{\frac{\Delta_{\epsilon}}{2}} \log(v)$:
\be
\delta_\epsilon \gamma_{\ell} = \delta_{\epsilon} \gamma_{\ell}^+ = - P_{\epsilon} \frac{\xi_{\Delta_{\phi},\Delta_{\phi}}^{\epsilon}}{\ell^{\Delta_{\epsilon}}} .
\ee
Finally, by adding or subtracting (\ref{eq:complex3}) from (\ref{eq:complex1}), we can project onto the even or odd spin uncharged double-twist operators. By matching terms of order $u^{\frac{\tau_{c}}{2}} \log(v)$ we then see that the existence of the charged scalar $c$ induces contributions of opposite sign to the even-spin and odd-spin anomalous dimensions:
\be\label{eq:oddu1shift}
\delta_{c} \gamma_{\ell^+} = - \delta_{c} \gamma_{\ell^-} = - P_c^+ \frac{\xi_{\Delta_{\phi},\Delta_{\phi}}^{c}}{\ell^{\Delta_{c}}} .
\ee

At first sight, the positive contributions to the anomalous dimensions of odd-spin operators in (\ref{eq:oddu1shift})  may look worrying in light of the Nachtmann theorem~\cite{Nachtmann:1973mr, Komargodski:2012ek} regarding convexity of the leading twist operators. However, it is important to note that the argument only applies to operators of {\it even spin} in reflection-positive OPEs, such as $\phi^{\dagger} \times \phi$.\footnote{In the notation of~\cite{Komargodski:2012ek}, the reason is that the amplitude $A(\nu,q^2) = \int d^d y e^{i q y} \<P|T \phi^{\dagger}(y)\phi(0)|P\>$ with $\nu \equiv 2 q\cdot P$ is no longer symmetric under $\nu \rightarrow -\nu$, so the moment $\mu_{\ell}(q^2)$ receives distinct contributions from both branch cuts in the complex $\nu$ plane. This gives $\mu_{\ell}(q^2) = \frac12 \int_0^1 dx x^{\ell-1} \left[ \text{Im} A(x,q^2) + (-1)^{\ell} \text{Im} A(-x,q^2)\right]$, with $x \equiv -q^2/\nu$, which is monotonic only for even $\ell$ after imposing the unitarity condition $\text{Im} A(x,q^2) > 0$. }

\subsection{$O(N)$}
\label{sec:SON}

In this section we will generalize the above discussion to the situation where the CFT has an $O(N)$ global symmetry.

\subsubsection{Fundamentals}

Let us first take $\phi_{i}$ to be in the fundamental representation of $O(N)$. This is the situation considered in the context of the numerical bootstrap in e.g.~\cite{Rattazzi:2010yc,Poland:2011ey,Vichi:2011ux,Kos:2013tga,Bae:2014hia,Nakayama:2014yia,Chester:2014gqa,Kos:2015mba}. We will start by rewriting the crossing conditions used in these works in a form that is suitable for our analysis.

Concretely, if we write the generic tensor structure of the 4-point function and switch $x_{2}\leftrightarrow x_{4}$ and $i_{2}\leftrightarrow i_{4}$, we obtain the condition
\begin{align}
\label{eq:SON4point}
x_{12}^{2\Delta_{\phi}} x_{34}^{2\Delta_{\phi}} \langle\phi_{i_{1}}(x_{1})\phi_{i_{2}}(x_{2})&\phi_{i_{3}}(x_{3})\phi_{i_{4}}(x_{4})\rangle  \nn\\
=\,\, &\delta_{i_{1}i_{2}}\delta_{i_{3}i_{4}}I_{s}(u,v)+\left(\delta_{i_{1}i_{4}}\delta_{i_{2}i_{3}}-\delta_{i_{1}i_{3}}\delta_{i_{2}i_{4}}\right)A_{s}(u,v) \nn\\
&+\left(\delta_{i_{1}i_{4}}\delta_{i_{2}i_{3}}+\delta_{i_{1}i_{3}}\delta_{i_{2}i_{4}}-\frac{2}{N}\delta_{i_{1}i_{2}}\delta_{i_{3}i_{4}}\right)S_{s}(u,v) \\
=\,\, &\left(\frac{u}{v} \right)^{\Delta_{\phi}} \Big( \delta_{i_{1}i_{4}}\delta_{i_{3}i_{2}}I_{t}(v,u)+\left(\delta_{i_{1}i_{2}}\delta_{i_{4}i_{3}}-\delta_{i_{1}i_{3}}\delta_{i_{2}i_{4}}\right)A_{t}(v,u)  \nn\\& +\left(\delta_{i_{1}i_{2}}\delta_{i_{4}i_{3}}+\delta_{i_{1}i_{3}}\delta_{i_{2}i_{4}}-\frac{2}{N}\delta_{i_{1}i_{4}}\delta_{i_{3}i_{2}}\right)S_{t}(v,u) \Big).
\end{align}
Solving for the functions in the t-channel expansion then gives
\be\label{eq:SONcrossing}
\left(\frac{u}{v} \right)^{\Delta_{\phi}} I_{t}(v,u) &=& \frac{1}{N} I_{s}(u,v)+\left(1-\frac{1}{N}\right)A_{s}(u,v)+\left(1+\frac{1}{N}-\frac{2}{N^2}\right)S_{s}(u,v),\nn\\
\left(\frac{u}{v} \right)^{\Delta_{\phi}} A_{t}(v,u) &=&\frac12 I_{s}(u,v)+ \frac12 A_{s}(u,v)-\frac12 \left(1+\frac{2}{N}\right)S_{s}(u,v),\nn\\
\left(\frac{u}{v} \right)^{\Delta_{\phi}}S_{t}(v,u) &=&\frac12 I_{s}(u,v)- \frac12 A_{s}(u,v)+ \frac12\left(1-\frac{2}{N}\right)S_{s}(u,v).
\ee

Focusing on the regime $u \ll v \ll 1$, in the $(12)-(34)$ channel we have contributions from the identity operator, singlet scalars $\epsilon$, symmetric tensor scalars $t_{ij}$, the $O(N)$ current $J_\mu$, and the stress tensor $T_{\mu\nu}$: 
\be
I_{s}(u,v)&\approx&1+P_{\epsilon} g_{\Delta_{\epsilon},0}(u,v)+P_T g_{d-2,2}(u,v),\nn\\ 
A_{s}(u,v)&\approx& P_J g_{d-2,1}(u,v),\nn\\ 
S_{s}(u,v)&\approx& P_t g_{\Delta_t,0}(u,v).
\ee
In all cases we define the coefficients $P_{\cO}$ by projecting the full contraction of the 3-point structures $\frac{1}{2^{\ell}} \lambda_{\phi_{i_1} \phi_{i_2} \cO_I} \lambda_{\cO^I \phi_{i_3} \phi_{i_4}}$ from the conformal OPE onto the tensor structures in (\ref{eq:SON4point}).

On the other hand, the $(14)-(32)$ channel has three types of double-twist operators in different $O(N)$ representations:
\begin{equation}
\cO_{\ell}^I=\phi_{i}\partial^{\ell}\phi_{i},\hspace{1em}\cO_{\ell}^A=\phi_{[i}\partial^{\ell}\phi_{j]},\hspace{1em}\cO_{\ell}^S=\phi_{(i}\partial^{\ell}\phi_{j)}-\frac{1}{N}\delta_{ij}\phi_{k}\partial^{\ell}\phi_{k},
\end{equation}
and the functions $I_t(u,v)$, $A_t(u,v)$, and $S_t(u,v)$ sum over these contributions using the cross-channel conformal blocks
\be
I_{t}(v,u) &\approx& \sum_{\ell^+} P_{\cO_{\ell}^I} g_{2\Delta_{\phi} + \gamma_{\ell}^I,\ell}(v,u), \nn\\
A_{t}(v,u) &\approx& \sum_{\ell^-} P_{\cO_{\ell}^A}g_{2\Delta_{\phi} + \gamma_{\ell}^A,\ell}(v,u),\nn\\
S_{t}(v,u) &\approx& \sum_{\ell^+} P_{\cO_{\ell}^S} g_{2\Delta_{\phi} + \gamma_{\ell}^S,\ell}(v,u),
\ee
where all of the coefficients are real and positive in unitary theories. 

Now by matching the identity contribution in (\ref{eq:SONcrossing}) as in the previous sections, we can easily read off the asymptotic behavior of the OPE coefficients to be
\be
\frac{N}{2} P_{\cO_{\ell}^I} = P_{\cO_{\ell}^A} = P_{\cO_{\ell}^S} = P_{\Delta_{\phi},\Delta_{\phi}}(\ell).
\ee
Similarly, by matching terms of order $u^{\frac{d-2}{2}} \log(v)$ in (\ref{eq:SONcrossing}) we obtain the corrections to the anomalous dimensions from stress-tensor exchange
\be
\delta_T \gamma_{\ell}^I=\delta_T \gamma_{\ell}^A=\delta_T \gamma_{\ell}^S=- \frac{d^2 \Delta_{\phi}^2}{4 (d-1)^2 C_{T}S_{d}^{2}} \frac{\xi_{\Delta_{\phi},\Delta_{\phi}}^{T}}{\ell^{d-2}}.
\ee
As above, we have inserted the value of $P_T$ as determined by the Ward identity (see Appendix~\ref{app:scalarOPE}). Again we see that the corrections due to stress-tensor exchange are universal and negative, which is consistent with a universal and attractive gravitational interaction in the bulk. The corrections due to the current exchange are
\be\label{eq:SONFundamentalAnomalousDimensionsJ}
\frac{1}{N-1} \delta_J \gamma_{\ell}^I =\delta_J \gamma_{\ell}^A=-\delta_J \gamma_{\ell}^S=- \frac{1}{2 C_{J}S_{d}^{2}} \frac{\xi_{\Delta_{\phi},\Delta_{\phi}}^{J}}{\ell^{d-2}}.
\ee

These shifts exhibit more structure. First, the signs of the corrections are determined by the representation of the composite operator. The singlet representation always has the largest negative anomalous dimension. This is consistent with our intuition about gauge interactions in the bulk, that two-particle states carrying the minimum charge are less energetic than other configurations. Another feature is that the ratios between different gauge binding energies are determined by the group structure and are independent of the dynamical details of the theory. 

For scalar exchange the corrections depend on the $O(N)$ representation of the scalar. Singlets give a universal contribution
\be
\delta_{\epsilon} \gamma_{\ell}^I=\delta_{\epsilon} \gamma_{\ell}^A=\delta_{\epsilon} \gamma_{\ell}^S=-P_{\epsilon} \frac{\xi_{\Delta_{\phi},\Delta_{\phi}}^{\epsilon}}{\ell^{\Delta_{\epsilon}}},
\ee
while symmetric tensor exchange gives corrections of either sign that depend on $N$
\be
\frac{N}{N^2+N-2} \delta_{t} \gamma_{\ell}^I=-\frac{N}{N+2}\delta_{t} \gamma_{\ell}^A=+\frac{N}{N-2}\delta_{t} \gamma_{\ell}^S=-P_t \frac{\xi_{\Delta_{\phi},\Delta_{\phi}}^{t}}{\ell^{\Delta_{t}}}.
\ee

Note that in the special case that $N=2$, these results reduce to the $U(1)$ case considered in the previous section after identifying $P_c = 2 P_t$. Similar to the discussion in~\cite{Poland:2011ey,Kos:2015mba}, the results are also valid in the special cases $N=3,4$, where the additional identifications and possible structures involving $\epsilon$-tensors do not lead to any modification of these results.

\subsubsection{Adjoints}\label{subsec:ONAdjoints}

Next let us consider the case of 4-point functions of $O(N)$ adjoints $\<AAAA\>$. The adjoint representation of $O(N)$ is the same as the $\frac{N(N-1)}{2}$-dimensional anti-symmetric representation $\tiny\yng(1,1)$, whose tensor product with itself admits the decomposition
\be\label{eq:adjointreps}
{\Yvcentermath1 \yng(1,1) \otimes \yng(1,1) = I \oplus \yng(1,1) \oplus \yng(2) \oplus \yng(2,2) \oplus \yng(2,1,1) \oplus \yng(1,1,1,1)}\,,
\ee
where all symmetrizations have traces removed. Double-twist operators $\cO_{\ell}^r$ in each of these representations appear in the $A \times A$ OPE, with spins restricted to be even/odd according to $(+,-,+,+,-,+)$ in the order of representations shown in~(\ref{eq:adjointreps}).

Again we can decompose the 4-point function into a sum over tensor structures
\be
x_{12}^{2\Delta_A} x_{34}^{2\Delta_A} \<A_{i_1 j_1}(x_1) A_{i_2 j_2}(x_2) A_{i_3 j_3}(x_3) A_{i_4 j_4}(x_4)\> = \sum_r (t^{A,r})_{i_1 i_2 i_3 i_4 j_1 j_2 j_3 j_4} G^{A,r}(u,v),
\ee
where $r$ runs over each possible representation. By expanding $G^{A,r}(u,v) = \sum P_{\cO^r} g_{\tau,\ell}(u,v)$ in conformal blocks, we can write out the crossing symmetry conditions relating the $(12)-(34)$ OPE to the $(14)-(32)$ OPE. We give the detailed form of the tensor structures and crossing equations in Appendix~\ref{app:structs}.

Applying the same logic, we first match the contributions of the identity operator, fixing the asymptotic behavior of the coefficients to be
\be
 P_{\cO_{\ell}^r} = P_{\Delta_A,\Delta_A}(\ell)\left( \frac{4}{N(N-1)}, \frac{1}{N-2}, \frac{1}{N-2}, \frac{2}{3}, 1, \frac{1}{3}\right) 
\ee
where again we show the representations in the same order as~(\ref{eq:adjointreps}).

Matching the stress-tensor block, we again see universal contributions to the anomalous dimensions
\be
\delta_T \gamma_{\ell}^r = -\frac{d^2 \Delta_{A}^2}{4 (d-1)^2 C_{T}S_{d}^{2}} \frac{\xi_{\Delta_{A},\Delta_{A}}^{T}}{\ell^{d-2}} \left(1,1,1,1,1,1\right),
\ee
for each representation $r$, but representation-dependent corrections from current exchange
\be\label{eq:SONAdjointAnomalousDimensionJ}
\delta_J \gamma_{\ell}^r = - \frac{1}{2C_{J}S_{d}^{2}} \frac{\xi_{\Delta_{A},\Delta_{A}}^{J}}{\ell^{d-2}} \left(2N-4, N-2, N-4,-2, 0, 4 \right).
\ee
We again notice that very similar structures show up as in the fundamental case. Note in particular that the fifth family of double trace states (with a ``hook"-like Young diagram) receive no anomalous dimensions at leading order in the large spin expansion. In a weakly-coupled bulk description, this result can be understood as a cancellation between the binding energy of the symmetrized and anti-symmetrized fundamental ``components" of the adjoint representation. 

From exchange of a scalar $\phi$ in representation $r'$, we obtain a set of shifts described by the matrix
\begin{align}
&\delta_{r'}\gamma_{\ell}^r = -P_{\phi^{r'}} \frac{\xi_{\Delta_{A},\Delta_{A}}^{\phi}}{\ell^{\Delta_{\phi}}} \times \nn\\
&\left( {\setlength\arraycolsep{1pt}\begin{array}{cccccc} 
1&1&1&1&1&1\\ 
2N-4&N-2&N-4&-2&0&4\\
\frac{2(N+2)(N-2)}{N} & \frac{(N+2)(N-4)}{N} & \frac{N^2-8}{N} & \frac{2N-8}{N} & -\frac{8}{N} & -\frac{4N+8}{N} \\ 
\frac{(N+2)(N+1)(N-3)}{2(N-1)} & - \frac{(N+2)(N+1)(N-3)}{2(N-1)(N-2)} & \frac{(N+1)(N-3)(N-4)}{2(N-1)(N-2)} & \frac{N^2-6N+11}{(N-1)(N-2)} & -\frac{(N+1)(N-4)}{(N-1)(N-2)} & \frac{(N+2)(N+1)}{(N-1)(N-2)} \\ 
\frac{(N+2)(N-3)}{2} & 0 & -\frac{2(N-3)}{(N-2)} & -\frac{N-4}{N-2} & 1 & -\frac{N+2}{N-2} \\
\frac{(N-2)(N-3)}{2} & N-3 & -(N-3) & 1 & -1 & 1
\end{array}}
\right),\nn\\
\end{align}
where e.g. the first row corresponds to exchanging a singlet, the second row corresponds to exchanging an anti-symmetric representation, etc. Note that in a 4-point function of identical scalars, $P_{\phi^{r}}=0$ for $r=\tiny\yng(1,1),\tiny\yng(2,1,1)$. But they may be nonzero when the external scalars are not identical.   

Finally, we mention a few special cases. For $N=4$, the 4-index anti-symmetric representation can be identified with an $SO(4)$ singlet that is odd under the $\mathbb{Z}_2$ subgroup of $O(4)$, while the hook representation can be identified with a $\mathbb{Z}_2$ odd symmetric tensor. We have checked that the results after these identifications are the same as the general results at $N=4$.\footnote{If the theory has $SO(N)$ instead of $O(N)$ symmetry, then for $N\le8$, the 4-point function may have additional structures containing $\epsilon$ tensors, but they only transform into themselves under crossing. So our crossing equations and results still apply. The $\epsilon$ structures will generate another set of crossing equations that may produce interesting constraints, but we will leave this for future study.} 
Also note that in (\ref{eq:SONAdjointAnomalousDimensionJ}), the anomalous dimensions of the aforementioned pairs become degenerate when $N=4$. For $N=2$ or $N=3$, the adjoint is equivalent to the singlet/fundamental and the results from the previous sections can be applied.

\subsubsection{Symmetric Tensors}
We can also repeat the analysis for 4-point functions of $O(N)$ symmetric tensors $\<SSSS\>$, where we have the tensor product
\be\label{eq:symreps}
{\Yvcentermath1 \yng(2) \otimes \yng(2) = I \oplus \yng(1,1) \oplus \yng(2) \oplus \yng(2,2) \oplus \yng(3,1) \oplus \yng(4)}.
\ee
In these cases the spins of the double-twist operators $\cO_{\ell}^r$ are restricted to $(+,-,+,+,-,+)$ in the order of representations shown in (\ref{eq:symreps}). We again decompose the 4-point function into tensor structures
\be
x_{12}^{2\Delta_S} x_{34}^{2\Delta_S} \<S_{i_1 j_1}(x_1) S_{i_2 j_2}(x_2) S_{i_3 j_3}(x_3) S_{i_4 j_4}(x_4)\> = \sum_r (t^{S,r})_{i_1 i_2 i_3 i_4 j_1 j_2 j_3 j_4} G^{S,r}(u,v),
\ee
and expand $G^{S,r}(u,v) = \sum P_{\cO^r} g_{\tau,\ell}(u,v)$ in conformal blocks, writing the structures and crossing relations explicitly in Appendix~\ref{app:structs}. 

Matching the identity contribution, the asymptotic behavior of the double-twist conformal block coefficients is determined to be
\be
P_{\cO_{\ell}^r} = P_{\Delta_S,\Delta_S}(\ell)\left( \frac{4}{(N+2)(N-1)}, \frac{1}{N+2}, \frac{N}{(N+4)(N-2)}, \frac{2}{3}, 1, \frac{1}{3}\right).
\ee
Matching the stress-tensor contribution gives universal contributions to the anomalous dimensions
\be
\delta_T \gamma_{\ell}^r = -\frac{d^2 \Delta_{\phi}^2}{4 (d-1)^2 C_{T}S_{d}^{2}} \frac{\xi_{\Delta_{S},\Delta_{S}}^T}{\ell^{d-2}} \left(1,1,1,1,1,1\right),
\ee
while the current gives
\be\label{eq:SONSymmetricAnomalousDimensionJ}
\delta_J \gamma_{\ell}^r = - \frac{1}{8C_{J}S_{d}^{2}} \frac{\xi_{\Delta_{S},\Delta_{S}}^{J}}{\ell^{d-2}} \left(2N, N+2, N,2, 0,-4 \right).
\ee

Finally, exchanging a scalar $\phi$ in a representation $r'$, gives contributions
\begin{align}
&\delta_{r'}\gamma_{\ell}^r = -P_{\phi^{r'}} \frac{\xi_{\Delta_{S},\Delta_{S}}^{\phi}}{\ell^{\Delta_{\phi}}} \times \nn\\
&\left( {\setlength\arraycolsep{1pt}\begin{array}{cccccc} 
1&1&1&1&1&1\\ 
2N&N+2&N&2&0&-4\\
\frac{2(N+4)(N-2)}{N} & \frac{(N+4)(N-2)}{N} & \frac{N^2+4N-24}{N} & -\frac{2N+8}{N} & -\frac{8}{N} & \frac{4N-8}{N} \\ 
\frac{N(N+1)(N-3)}{2(N-1)} & \frac{(N+1)(N-3)}{2(N-1)} & -\frac{N(N+1)(N-3)}{2(N-1)(N-2)} & \frac{N^2-2N+3}{(N-1)(N-2)} & -\frac{N(N-3)}{(N-1)(N-2)} & \frac{N-3}{N-1} \\ 
\frac{(N+4)(N+1)(N-2)}{2(N+2)} & 0 & -\frac{2(N+1)}{(N+2)} & -\frac{N+4}{N+2} & 1 & -\frac{N-2}{N+2} \\
\frac{N(N+6)(N+1)}{2(N+2)} & -\frac{(N+6)(N+1)}{(N+2)} & \frac{N(N+6)(N+1)}{(N+4)(N+2)} & \frac{N+6}{N+2} & -\frac{N(N+6)}{(N+4)(N+2)} & \frac{N(N-2)}{(N+4)(N+2)}
\end{array}}
\right).\nn\\
\end{align}
Note that in a 4-point function of identical scalars, $P_{\phi^{r}}=0$ for $r=\tiny\yng(1,1),\tiny\yng(3,1)$. But they may be nonzero when the external scalars are not identical.   

\subsection{$SU(N)$}
\label{sec:SUN}

Next we will study the double-twist asymptotics for CFTs with scalars charged under an $SU(N)$ global symmetry, focusing on the most common cases of fundamentals and adjoints.

\subsubsection{Fundamentals}

To begin, let us consider a 4-point function containing $SU(N)$ fundamentals and their conjugates, which can be decomposed in singlet and adjoint contributions in either the $(12)-(34)$ channel or the $(14)-(32)$ channel as
\begin{align}
\label{eq:SUNFundamentalCrossing}
x_{12}^{2\Delta_{\phi}} x_{34}^{2\Delta_{\phi}} \langle\phi_{i_{1}}(x_1)\phi^{\dagger i_{2}}(x_2)&\phi_{i_{3}}(x_3)\phi^{\dagger i_{4}}(x_4)\rangle \nn\\
&= \delta_{i_{1}}^{i_{2}}\delta_{i_{3}}^{i_{4}}I_{s}(u,v)+\left(\delta_{i_{1}}^{i_{4}}\delta_{i_{3}}^{i_{2}}-\frac{1}{N}\delta_{i_{1}}^{i_{2}}\delta_{i_{3}}^{i_{4}}\right)Adj_{s}(u,v) \nn\\
&= \left(\frac{u}{v}\right)^{\Delta_{\phi}} \left( \delta_{i_{1}}^{i_{4}}\delta_{i_{3}}^{i_{2}}I_{t}(v,u)+\left(\delta_{i_{1}}^{i_{2}}\delta_{i_{3}}^{i_{4}}-\frac{1}{N}\delta_{i_{1}}^{i_{4}}\delta_{i_{3}}^{i_{2}}\right)Adj_{t}(v,u)\right). \nn\\
\end{align}
Solving for the t-channel contributions gives the crossing equations
\be\label{eq:iad1}
\left(\frac{u}{v}\right)^{\Delta_{\phi}} I_t(v,u) &=& \frac{1}{N} I_s(u,v) + \left(1-\frac{1}{N^2}\right) Adj_s(u,v), \nn\\
\left(\frac{u}{v}\right)^{\Delta_{\phi}} Adj_t(v,u) &=& I_s(u,v) - \frac{1}{N} Adj_s(u,v).
\ee

In the lightcone limit, we can approximate
\be\label{eq:sunlightcone}
I_s(u,v) \approx 1 + P_{\epsilon} g_{\Delta_{\epsilon},0} + P_T g_{d-2,2}(u,v), \nn\\
Adj_s(u,v) \approx P_a g_{\Delta_{a},0}(u,v) + P_J g_{d-2,1}(u,v),
\ee
where we have included the contributions of the lowest-twist singlet scalar $\epsilon$ and adjoint scalar $a$.

On the other hand, the t-channel functions sum over the contributions of the double-twist operators $\cO_{\ell}^I \sim \phi_i \partial^{\ell} \phi^{\dagger i}$ and $\cO_{\ell}^{Adj} \sim \phi_i \partial^{\ell} \phi^{\dagger j}$:
\be
I_t(v,u) &\approx& \sum_{\ell^+} P_{\cO_{\ell^+}^I} g_{2\Delta_{\phi} + \gamma_{\ell^+}^I}(v,u)+\sum_{\ell^-} P_{\cO_{\ell^-}^I} g_{2\Delta_{\phi} + \gamma_{\ell^-}^I}(v,u),\nn\\
Adj_t(v,u) &\approx& \sum_{\ell^+} P_{\cO_{\ell^+}^{Adj}} g_{2\Delta_{\phi} + \gamma_{\ell^+}^{Adj}}(v,u)+\sum_{\ell^-} P_{\cO_{\ell^-}^{Adj}} g_{2\Delta_{\phi} + \gamma_{\ell^-}^{Adj}}(v,u), \label{eqn:SU(N)_OPE}
\ee
where for clarity we have written separately the even-spin and odd-spin contributions.

Let us finally consider the constraint connecting the symmetric and anti-symmetric tensors in $\phi \times \phi$ to the singlet and adjoint operators in $\phi \times \phi^{\dagger}$:
\begin{align}
x_{12}^{2\Delta_{\phi}} x_{34}^{2\Delta_{\phi}} \<\phi_{i_{1}}(x_1)\phi_{i_{2}}(x_2)&\phi^{\dagger i_{3}}(x_3)\phi^{\dagger i_{4}}(x_4)\> \nn\\ &= \left(\delta_{i_{1}}^{i_{3}}\delta_{i_{2}}^{i_{4}}+\delta_{i_{1}}^{i_{4}}\delta_{i_{2}}^{i_{3}}\right)S(u,v)+\left(\delta_{i_{1}}^{i_{4}}\delta_{i_{2}}^{i_{3}}-\delta_{i_{1}}^{i_{3}}\delta_{i_{2}}^{i_{4}}\right)A(u,v) \nn\\
&= \left(\frac{u}{v}\right)^{\Delta_{\phi}} \left( \delta_{i_{1}}^{i_{4}}\delta_{i_{2}}^{i_{3}}\tilde{I}(v,u)+\left(\delta_{i_{1}}^{i_{3}}\delta_{i_{2}}^{i_{4}}-\frac{1}{N}\delta_{i_{1}}^{i_{4}}\delta_{i_{2}}^{i_{3}}\right)\tilde{Adj}(v,u)\right), \nn\\
\end{align}
giving the conditions
\be\label{eq:sa1}
 \left(\frac{u}{v}\right)^{\Delta_{\phi}}\tilde{I}(v,u) &=& \left(1+\frac{1}{N}\right) S(u,v) + \left(1-\frac{1}{N}\right) A(u,v), \nn\\
  \left(\frac{u}{v}\right)^{\Delta_{\phi}}\tilde{Adj}(v,u) &=& S(u,v) - A(u,v),
\ee
or equivalently on switching $u \leftrightarrow v$
\be\label{eq:sa2}
 \left(\frac{u}{v}\right)^{\Delta_{\phi}}S(v,u) &=& \frac{1}{2} \tilde{I}(u,v) + \frac12 \left(1-\frac{1}{N}\right) \tilde{Adj}(u,v), \nn\\
  \left(\frac{u}{v}\right)^{\Delta_{\phi}}A(v,u) &=& \frac{1}{2} \tilde{I}(u,v) - \frac12 \left(1+\frac{1}{N}\right) \tilde{Adj}(u,v).
\ee
Both (\ref{eq:sa1}) and (\ref{eq:sa2}) give interesting information in the $u \ll v \ll 1$ limit.

In particular, the RHS of (\ref{eq:sa1}) at small $u$ probes the low-twist scalar symmetric tensors $s$, giving
\be
S(u,v) \approx P_s^+ g_{\Delta_s,0}(u,v),
\ee
while the RHS of (\ref{eq:sa2}) probes the low-twist singlets and adjoints, with the current having an opposite relative sign compared to (\ref{eq:sunlightcone}),
\be
\tilde{I}(u,v) \approx 1 + P_{\epsilon} g_{\Delta_{\epsilon},0} + P_T g_{d-2,2}(u,v), \nn\\
\tilde{Adj}(u,v) \approx P_a g_{\Delta_{a},0}(u,v) - P_J g_{d-2,1}(u,v).
\ee

On the other hand, the LHS of (\ref{eq:sa1}) distinguishes between the even- and odd-spin uncharged double-twist operators
\be
\tilde{I}(v,u) &\approx& \sum_{\ell^+} P_{\cO_{\ell^+}^I} g_{2\Delta_{\phi} + \gamma_{\ell^+}^I}(v,u)-\sum_{\ell^-} P_{\cO_{\ell^-}^I} g_{2\Delta_{\phi} + \gamma_{\ell^-}^I}(v,u),\nn\\
\tilde{Adj}(v,u) &\approx& \sum_{\ell^+} P_{\cO_{\ell^+}^{Adj}} g_{2\Delta_{\phi} + \gamma_{\ell^+}^{Adj}}(v,u)-\sum_{\ell^-} P_{\cO_{\ell^-}^{Adj}} g_{2\Delta_{\phi} + \gamma_{\ell^-}^{Adj}}(v,u),
\ee
while the LHS of (\ref{eq:sa2}) probes the symmetric and anti-symmetric tensor double-twist operators $\cO^S_{\ell^+} \sim \phi_{(i} \partial^{\ell} \phi_{j)}$ and $\cO^A_{\ell^-} \sim \phi_{[i} \partial^{\ell} \phi_{j]}$,
\be
S(v,u) &\approx& \sum_{\ell^+} P^+_{\cO_{\ell^+}^S} g_{2\Delta_{\phi} + \gamma_{\ell^+}^S}(v,u),\nn\\
A(v,u) &\approx& -\sum_{\ell^-} P^+_{\cO_{\ell^-}^{A}} g_{2\Delta_{\phi} + \gamma_{\ell^-}^{A}}(v,u),
\ee

Matching the identity operator on the RHS of (\ref{eq:iad1}) and (\ref{eq:sa2}) gives the asymptotic behavior of the OPE coefficients
\be
P_{\cO^{Adj}_{\ell}} = N P_{\cO^{I}_{\ell}} = P_{\cO^{S}_{\ell^+}} = P^+_{\cO^{A}_{\ell^-}} = P_{\Delta_{\phi},\Delta_{\phi}}(\ell),
\ee
while matching the $\log(v)$ singularities as expected gives universal negative shifts in the anomalous dimensions from stress-tensor exchange and singlet scalar exchange
\be
\delta_T \gamma_{\ell}^r =- \frac{d^2 \Delta_{\phi}^2}{4 (d-1)^2 C_{T}S_{d}^{2}} \frac{\xi_{\Delta_{\phi},\Delta_{\phi}}^T}{\ell^{d-2}},\,\,\, \delta_{\epsilon} \gamma_{\ell}^r = -P_{\epsilon} \frac{\xi_{\Delta_{\phi},\Delta_{\phi}}^{\epsilon}}{\ell^{\Delta_{\epsilon}}},
\ee
shifts from current and adjoint scalar exchange of a similar form given by
\be\label{eq:SUNFundamentalAnomalousDimensionJ}
\frac{N}{N^2-1} \delta_J \gamma_{\ell}^{I} &=& -N \delta_J \gamma_{\ell}^{Adj} = - \frac{N}{N-1} \delta_J \gamma_{\ell^+}^S = \frac{N}{N+1} \delta_J \gamma_{\ell^-}^A = - \frac{1}{2 C_{J}S_{d}^{2}} \frac{\xi_{\Delta_{\phi},\Delta_{\phi}}^J}{\ell^{d-2}},\\
\frac{N}{N^2-1} \delta_a \gamma_{\ell}^{I} &=& -N \delta_a \gamma_{\ell}^{Adj} = - \frac{N}{N-1} \delta_a \gamma_{\ell^+}^S = \frac{N}{N+1} \delta_a \gamma_{\ell^-}^A = - P_a \frac{\xi_{\Delta_{\phi},\Delta_{\phi}}^a}{\ell^{\Delta_a}},
\ee
and shifts to $\gamma^I_{\ell}$ and $\gamma^{Adj}_{\ell}$ from symmetric-tensor scalar exchange that gives opposite sign contributions to even and odd spins
\be
\frac{1}{N+1} \delta_{s} \gamma_{\ell^+}^I = - \frac{1}{N+1} \delta_{s} \gamma_{\ell^-}^I = \delta_s \gamma_{\ell^+}^{Adj} = - \delta_s \gamma_{\ell^-}^{Adj} = - P^+_s \frac{\xi_{\Delta_{\phi},\Delta_{\phi}}^s}{\ell^{\Delta_s}}.
\ee

\subsubsection{Adjoints}
Finally we consider the situation of 4-point functions of $SU(N)$ adjoints $\Phi_{i}^j$. This case was considered in the numerical 4D bootstrap in~\cite{Berkooz:2014yda}. Following the notation of~\cite{Berkooz:2014yda}, the tensor product contains 7 irreducible representations
\be\label{eq:adjreps}
Adj \otimes Adj = I \oplus Adj_a \oplus Adj_s \oplus \left((S,\bar{A})_a \oplus (A,\bar{S})_a \right) \oplus (A,\bar{A})_s \oplus (S,\bar{S})_s,
\ee
where the subscript $s$ or $a$ denotes whether the representation is in the symmetric or anti-symmetric product of the adjoints and we group together $\left((S,\bar{A})_a \oplus (A,\bar{S})_a \right) $ because they are conjugates and will have identical dimensions and OPE coefficients. In the case of identical adjoints, the representations on the RHS of (\ref{eq:adjreps}) can appear with spins $(+,-,+,-,+,+)$.

As before, we decompose the 4-point function into tensor structures
\be\label{eq:adj4pt}
x_{12}^{2\Delta_{\Phi}} x_{34}^{2\Delta_{\Phi}} \<\Phi_{i_1}^{j_1}(x_1) \Phi_{i_2}^{j_2}(x_2) \Phi_{i_3}^{j_3}(x_3) \Phi_{i_4}^{j_4}(x_4)\> = \sum_r (t^{\Phi,r})_{i_1 i_2 i_3 i_4}^{j_1 j_2 j_3 j_4} G^{\Phi,r}(u,v),
\ee
and expand $G^{\Phi,r}(u,v) = \sum P_{\cO^r} g_{\tau,\ell}(u,v)$ in conformal blocks, writing the structures and crossing relations explicitly in Appendix~\ref{app:structs}.

Matching the identity gives the asymptotic coefficients of the double-twist operators
\be
P_{\cO_{\ell}^r} = P_{\Delta_{\Phi},\Delta_{\Phi}}(\ell) \left(\frac{2}{(N+1)(N-1)},\frac{1}{N},\frac{N}{(N+2)(N-2)},1,\frac12,\frac12\right),
\ee
matching the stress-tensor gives a universal shift
\be
\delta_T \gamma_{\ell}^r &=& -\frac{d^2 \Delta_{\phi}^2}{4 (d-1)^2 C_{T}S_{d}^{2}} \frac{\xi_{\Delta_{\Phi},\Delta_{\Phi}}^T}{\ell^{d-2}} \left(1,1,1,1,1,1\right),
\ee
matching the currents gives the shifts
\be\label{eq:SUNAdjointAnomalousDimensionJ}
\delta_J \gamma_{\ell}^r &=& -\frac{1}{2C_{J}S_{d}^{2}} \frac{\xi_{\Delta_{\Phi},\Delta_{\Phi}}^J}{\ell^{d-2}} \left(2N,N,N,0,2,-2\right),
\ee
and matching contributions from a scalar $\phi$ in representation $r'$ gives the matrix
\begin{align}
&\delta_{r'}\gamma_{\ell}^r = -P_{\phi^{r'}} \frac{\xi_{\Delta_{\Phi},\Delta_{\Phi}}^{\phi}}{\ell^{\Delta_{\phi}}} \times \nn\\
&\left( {\setlength\arraycolsep{1pt}\begin{array}{cccccc} 
1&1&1&1&1&1\\ 
2N&N&N&0&2&-2\\
\frac{2(N+2)(N-2)}{N} & \frac{(N+2)(N-2)}{N} & \frac{N^2-12}{N} & -\frac{4}{N} & -\frac{2N+4}{N} & \frac{2N-4}{N} \\ 
(N+2)(N-2) & 0 & -2 & 1 & -\frac{N+2}{N} &-\frac{N-2}{N} \\ 
\frac{N^2(N-3)}{(N-1)} & \frac{N(N-3)}{N-1} & -\frac{N^2(N-3)}{(N-2)(N-1)} & -\frac{N(N-3)}{(N-2)(N-1)} & \frac{N^2-N+2}{(N-2)(N-1)} & \frac{N-3}{N-1} \\
\frac{N^2(N+3)}{(N+1)} & -\frac{N(N+3)}{(N+1)} & \frac{N^2(N+3)}{(N+2)(N+1)} & -\frac{N(N+3)}{(N+2)(N+1)} & \frac{N+3}{N+1} & \frac{N^2+N+2}{(N+2)(N+1)}
\end{array}}
\right).\nn\\
\end{align}
Note that in a 4-point function of identical scalars, $P_{\phi^{r}}=0$ for $r=Adj_a$ or $(S,\bar{A})_a \oplus (A,\bar{S})_a $. But they may be nonzero when the external scalars are not identical.  
 
In the case of $SU(3)$ the $(A,\bar{A})_s$ representation does not exist but otherwise the results apply with $N=3$. In the case of $SU(2)$ only the $(S,\bar{S})_s$, $Adj_a$, and trivial representations exist, but the results for these operators are also correct after setting $N=2$.\footnote{Comparing to the $O(3)$ case, there is an apparent factor of 2 arising from different normalization of the generator and the current central charge. In our conventions for $SU(2)$, $f_{abc}=\sqrt{2}\epsilon_{abc}$ and $C_J^{SU(2)}=2 C_J^{O(3)}$. See Appendix~\ref{app:scalarOPE} for more details.}

\section{Applications}
\label{sec:applications}

\subsection{4D $\mathcal{N}=1$ SQCD}
\label{subsec:SQCD}

In this section, we apply our results to interacting 4D
$\mathcal{N}=1$ superconformal field theories (SCFTs). The supersymmetric unitarity
bound for real operators, $\Delta \ge \ell + 2$, forbids real scalars with
twist lower than that of conserved currents. For interacting SCFTs,
a real scalar with $\Delta=2$ belongs to a supermultiplet containing
a conserved current. 

For concreteness, we consider $\mathcal{N}=1$
SCFTs with an $SU(F)$ global symmetry and focus on the 4-point function $\langle\tilde{\phi}_{a_{1}}\tilde{\phi}_{a_{2}}\tilde{\phi}_{a_{3}}\tilde{\phi}_{a_{4}}\rangle$,
where $\tilde{\phi}_{a}$ is the canonically-normalized lowest component
of the $SU(F)$ current multiplet and $a$ is an adjoint index. For the expansion of this 4-point function, we have
\begin{equation}
P_{T}=\frac{1}{90c},\hspace{1em}P_{J}=\frac{1}{6\tau},\hspace{1em}P_{\phi^{Adj_{s}}}=\frac{1}{2}\frac{\kappa^{2}}{\tau^{3}},\hspace{1em}P_{\phi^{I}}=\frac{2\kappa_{1}^{2}}{\tau^{2}\tau_{1}},
\end{equation}
where $c=\frac{C_{T}\pi^{4}}{40}$ is the coefficient of the Weyl anomaly
in $\langle T_{\mu}^{\mu}\rangle=\frac{c}{16\pi^{2}}(Weyl)^{2}-\frac{a}{16\pi^{2}}(Euler)^{2}$.
$C_{J}=\frac{3\tau}{4\pi^{4}}$ is the flavor central charge for $SU(F)$. In SCFTs
$\tau$ is related to the $SU(F)^{2}U(1)_{R}$ anomaly. $\tau_{1}$
is the similar quantity for a $U(1)$ current contained in the multiplet denoted as $\phi^I$. $\kappa$ and $\kappa_1$ are the coefficients of the $SU(F)^{3}$ and $SU(F)^{2}U(1)$ anomalies, respectively.

Let us now consider the specific case of $\mathcal{N}=1$ SQCD, which has an $SU(F)\times SU(F)\times U(1)_{B} \times U(1)_{R}$ global symmetry together with an $SU(N)$ gauge symmetry.  The theory flows to an interacting conformal fixed point when $\frac{3}{2}N<F<3N$~\cite{Seiberg:1994pq}.
The relevant coefficients can be computed by anomaly matching and are given by:
\begin{table}[H]
\centering{}%
\begin{tabular}{|c|c|c|c|c|}
\hline 
$\tau$ & $\tau_{1}$ & $\kappa$ & $\kappa_{1}$ & $c$\tabularnewline
\hline 
\hline 
$3\frac{N^{2}}{F}$ & $6N^{2}$ & $N$ & $N$ & $\frac{1}{16}\left(7N^{2}-9\frac{N^{4}}{F^{2}}-2\right)$\tabularnewline
\hline 
\end{tabular}\caption{The central charges and anomaly coefficients of $\mathcal{N}=1$ SQCD.}
\end{table}
Using these OPE coefficients, we can work out the anomalous dimensions of large spin double-twist operators $\tilde{\phi}_a \partial^\ell \tilde{\phi}_b$ at leading order in $1/\ell^2$, where $\tilde{\phi}_{a,b}$ are the lowest components of the current multiplet corresponding to one of the $SU(F)$ flavor symmetries. For $F=3$, we can only have $N=2$ on the boundary of the conformal window where the magnetic theory is free, so our analysis does not apply. For $F\ge4$, the result is 
\begin{align}
\gamma^{1} & =-\frac{2}{27}\frac{1}{\ell^{2}}\left(\frac{144F^{2}}{7N^{2}F^{2}-9N^{4}-2F^{2}}+\frac{18N^{2}F^{2}+F^{4}-3F^{2}}{N^{4}}\right),\\
\gamma^{2} & =-\frac{2}{27}\frac{1}{\ell^{2}}\left(\frac{144F^{2}}{7N^{2}F^{2}-9N^{4}-2F^{2}}+\frac{9N^{2}F^{2}+\frac{1}{2}F^{4}-F^{2}}{N^{4}}\right),\\
\gamma^{3} & =-\frac{2}{27}\frac{1}{\ell^{2}}\left(\frac{144F^{2}}{7N^{2}F^{2}-9N^{4}-2F^{2}}+\frac{9N^{2}F^{2}+\frac{1}{2}F^{4}-5F^{2}}{N^{4}}\right),\\
\gamma^{4} & =-\frac{2}{27\ell^{2}}\left(\frac{144F^{2}}{7N^{2}F^{2}-9N^{4}-2F^{2}}-\frac{F^{2}}{N^{4}}\right),\\
\gamma^{5} & =-\frac{2}{27}\frac{1}{\ell^{2}}\left(\frac{144F^{2}}{7N^{2}F^{2}-9N^{4}-2F^{2}}+\frac{18N^{2}F-F^{3}-F^{2}}{N^{4}}\right),\\
\gamma^{6} & =-\frac{2}{27}\frac{1}{\ell^{2}}\left(\frac{144F^{2}}{7N^{2}F^{2}-9N^{4}-2F^{2}}-\frac{18N^{2}F-F^{3}+F^{2}}{N^{4}}\right).
\end{align}
In these expressions the index $r$ of $\gamma^r$ labels the representation of the double-twist operator under $SU(F)$ as given in (\ref{eq:adjreps}). 

Note that the first term in these results is from the stress tensor exchange and is the same across different representations. 
Going to the Veneziano limit with $\ell \gg N,F\gg1$ and $F/N$ fixed to be in the conformal window, we have at leading order
\begin{align}
\gamma^{1} & =2\gamma^{2}=2\gamma^{3}=-\frac{2}{27}\frac{1}{\ell^{2}}\left[\left(\frac{F}{N}\right)^{4}+18\left(\frac{F}{N}\right)^{2}\right]+\mathcal{O}\left(\frac{1}{N^{2}}\right),\\
\gamma^{4} & =-\frac{2}{27}\frac{1}{\ell^{2}}\frac{1}{N^{2}}\left[\frac{144\left(\frac{F}{N}\right)^{2}}{7\left(\frac{F}{N}\right)^{2}-9}-\left(\frac{F}{N}\right)^2\right]+\mathcal{O}\left(\frac{1}{N^{4}}\right),\\
\gamma^{5} & =-\gamma^{6}=-\frac{2}{27}\frac{1}{\ell^{2}}\frac{1}{N}\left[18\left(\frac{F}{N}\right)-\left(\frac{F}{N}\right)^{3}\right]+\mathcal{O}\left(\frac{1}{N^{3}}\right).
\end{align}

Here we see that the anomalous dimensions of the singlet and adjoint representations ($r=1,2,3$) have no additional suppression at large $N$ while the anomalous dimensions of the remaining representations fall off like $1/N$ or $1/N^2$. This is reflecting the fact that the former states correspond to ``generalized single trace" operators with neighboring color and flavor indices contracted, while the latter representations describe ``generalized double trace" operators. See e.g.~\cite{Poland:2011kg} for a discussion of this large-$N$ counting in the Veneziano limit of SQCD. The existence of these ``generalized single trace" operators is still consistent with large-$N$ factorization because their OPE coefficients become suppressed at large $N$.

\subsection{3D $O(N)$ Vector Models}

Our analysis can also be applied to the 3D $O(N)$ vector models in the regime $\ell \gg N$ (for $N \gg \ell$ approximate higher spin currents must also be included, as we discuss in the next subsection). If we consider 4-point functions of the $O(N)$ fundamental $\phi^i$, then our analysis gives
\be
\gamma^I_{\ell} &=& - \left(\frac{9 \Delta_{\phi}^2}{2^{8} C_T\pi^{2}} \xi_{\Delta_{\phi},\Delta_{\phi}}^T + \frac{N-1}{32C_J\pi^{2}} \xi_{\Delta_{\phi},\Delta_{\phi}}^J\right) \frac{1}{\ell} - P_{\epsilon} \frac{\xi_{\Delta_{\phi},\Delta_{\phi}}^{\epsilon}}{\ell^{\Delta_{\epsilon}}} - \frac{N^2-N+2}{N} P_t \frac{\xi_{\Delta_{\phi},\Delta_{\phi}}^{t}}{\ell^{\Delta_{t}}} + \ldots,\nn\\
\gamma^A_{\ell} &=& - \left(\frac{9 \Delta_{\phi}^2}{2^{8} C_T\pi^{2}} \xi_{\Delta_{\phi},\Delta_{\phi}}^T + \frac{1}{32C_J\pi^{2}} \xi_{\Delta_{\phi},\Delta_{\phi}}^J\right) \frac{1}{\ell} - P_{\epsilon} \frac{\xi_{\Delta_{\phi},\Delta_{\phi}}^{\epsilon}}{\ell^{\Delta_{\epsilon}}} + \frac{N+2}{N} P_t \frac{\xi_{\Delta_{\phi},\Delta_{\phi}}^{t}}{\ell^{\Delta_{t}}} + \ldots,\nn\\
\gamma^S_{\ell} &=& - \left(\frac{9 \Delta_{\phi}^2}{2^{8} C_T\pi^{2}} \xi_{\Delta_{\phi},\Delta_{\phi}}^T - \frac{1}{32C_J\pi^{2}} \xi_{\Delta_{\phi},\Delta_{\phi}}^J\right) \frac{1}{\ell} - P_{\epsilon} \frac{\xi_{\Delta_{\phi},\Delta_{\phi}}^{\epsilon}}{\ell^{\Delta_{\epsilon}}} - \frac{N-2}{N} P_t \frac{\xi_{\Delta_{\phi},\Delta_{\phi}}^{t}}{\ell^{\Delta_{t}}} + \ldots,
\ee
where we included the stress tensor, $O(N)$ current, and the leading scalar singlet and $O(N)$ symmetric tensor.

In Table~\ref{tab:on} we summarize some numerical data for the most interesting cases $N=2,3$. In our normalization, the free field theory values are $C_T^{\text{free}} = \frac{3}{32\pi^{2}}$ and $C_J^{\text{free}} = \frac{1}{8\pi^{2}}$. Unfortunately, we are not aware of determinations of the coefficients $P_{\epsilon}$ and $P_{t}$, but it is likely that they can be extracted from future conformal bootstrap studies of the $O(N)$ vector models. 
\begin{table}[H]
\centering{}%
\begin{tabular}{|c|c|c|c|c|c|}
\hline 
$$ & $C_T/N C^{\text{free}}_T$ & $C_J/C^{\text{free}}_J$ & $\Delta_{\phi}$ & $\Delta_{\epsilon}$ & $\Delta_{t}$ \tabularnewline
\hline 
$O(2)$ & 0.94365(13)~\cite{Kos:2013tga} & 0.9050(16)~\cite{Kos:2015mba}  & 0.51905(10)~\cite{Campostrini:2006ms}  & 1.51124(22)~\cite{Campostrini:2006ms}  & 1.237(4)~\cite{Calabrese:2004ca} \tabularnewline
\hline
$O(3)$ & 0.94418(43)~\cite{Kos:2013tga} & 0.9065(27)~\cite{Kos:2015mba} &  0.51875(25)~\cite{Campostrini:2002ky} & 1.5939(10)~\cite{Campostrini:2002ky}  & 1.211(3)~\cite{Calabrese:2004ca} \tabularnewline
\hline 
\end{tabular}\caption{Numerical data for the $O(2)$ and $O(3)$ vector models.}
\label{tab:on}\end{table}

Plugging in numbers, we see that for the $O(2)$ model we have the corrections:
\be
\gamma^I_{\ell} &\simeq & -\frac{0.00310}{\ell} - P_t \frac{0.04389}{\ell^{1.237}} - P_{\epsilon} \frac{0.13388}{\ell^{1.5112}},\nn\\
\gamma^A_{\ell} &\simeq & -\frac{0.00310}{\ell} + P_t \frac{0.04389}{\ell^{1.237}} - P_{\epsilon} \frac{0.13388}{\ell^{1.5112}},\nn\\
\gamma^S_{\ell} &\simeq & -\frac{0.00005}{\ell} - P_{\epsilon} \frac{0.13388}{\ell^{1.5112}},
\ee
while for the $O(3)$ model we have:
\be
\gamma^I_{\ell} &\simeq & -\frac{0.00396}{\ell} - P_t \frac{0.07423}{\ell^{1.211}} - P_{\epsilon} \frac{0.18524}{\ell^{1.5939}},\nn\\
\gamma^A_{\ell} &\simeq & -\frac{0.00249}{\ell} + P_t \frac{0.02749}{\ell^{1.211}} - P_{\epsilon} \frac{0.18524}{\ell^{1.5939}},\nn\\
\gamma^S_{\ell} &\simeq & +\frac{0.00046}{\ell} - P_t \frac{0.00550}{\ell^{1.211}} - P_{\epsilon} \frac{0.18524}{\ell^{1.5939}}.
\ee
Interestingly, the twist 1 contribution to $\gamma^S_{\ell}$ appears to change sign between the $O(2)$ and the $O(3)$ model. On the other hand, the coefficients of the twist 1 terms are likely somewhat suppressed relative to the scalar contributions (assuming $\mathcal{O}(1)$ OPE coefficients), so at moderate values of $\ell$ we expect the latter to dominate. 

In the regime of moderate $\ell$ we also expect the contributions of higher spin operators of minimal twist to become more important. In~\cite{Alday:2015ota} it was seen that these will lead to corrections of the form $\delta \gamma_{\ell} \sim \frac{f(\log \ell)}{\ell}$ for some function $f(\log\ell)$ that could be computed in an expansion in the limit of approximate higher spin symmetry. In our context including only the twist 1 contributions, as above, corresponds to taking $f(\log\ell) \rightarrow f(\infty)$. It would be interesting to better understand the form of these higher spin corrections.

\subsection{CFTs with a Large Global Symmetry}
\label{subsec:largeN}

A generic feature of our results (\ref{eq:SONFundamentalAnomalousDimensionsJ}, \ref{eq:SONAdjointAnomalousDimensionJ}, \ref{eq:SONSymmetricAnomalousDimensionJ}, \ref{eq:SUNFundamentalAnomalousDimensionJ}, \ref{eq:SUNAdjointAnomalousDimensionJ}) is that the anomalous dimensions of certain double-twist operators grow with the size $N$ of the global symmetry group $O(N)$ or $SU(N)$. This implies that our results are only valid when $\ell\gg N$ unless $C_J$ grows fast enough to cancel this effect. 
Otherwise, some of these anomalous dimensions would violate the unitarity bounds and/or cannot be used as perturbative parameters. 

It is extremely interesting to understand what happens when $\ell\sim N$, but this regime is subtle to work with because the small $u$ expansion is not clearly separated from the large $N$ expansion. 
In this section, we focus instead on the opposite regime with $\ell\ll N$, where $N$ is sent to infinity. 
What we will see is that, under a set of assumptions, crossing symmetry requires the existence of an infinite number of conserved higher spin currents at $N\rightarrow\infty$ and that scalars with twist smaller than $d-2$ cannot appear in the OPE in this limit. 

To be concrete, we will demonstrate this idea in a unitary CFT with a $SU(N)$ global symmetry. 
We study the 4-point function of fundamental and antifundamental scalars, but the same argument applies to all cases where the anomalous dimensions at $\ell\gg N$ grow with $N$.
The first assumption is that there exist scalar operators transforming under the global symmetry. 
This is not essential since similar behavior is seen when studying the 4-point function of conserved currents \cite{CurrentPaper:2015}.
The second assumption is that 4-point functions of the canonically normalized operators have well-defined limits when $N\rightarrow\infty$. 
In the $SU(N)$ 4-point function we are considering, this means that $I_{s}(u,v)$ and $Adj_{s}(u,v)$ do not blow up as $N\rightarrow\infty$. 
This for example implies that $N C_{T}\rightarrow\infty$ as $N\rightarrow\infty$, which is natural because $C_{T}$ is generically expected to grow as a function of $N$. 
Finally, the key assumption we need is $C_{J}\rightarrow\mathcal{O}(1)$  as $N\rightarrow\infty$. That is, we will assume $C_{J}$ does not grow with $N$ for $N$ large.

For free field theories these assumptions hold if we only have a finite number of scalars and fermions in the fundamental representation of the flavor symmetry group. 
In general, the contribution of a free scalar or fermion in representation $r$ of the flavor symmetry group to $C_{J}$ is of order $C(r)$, or the index of the representation \cite{Osborn:1993cr}. 
The index is defined as $Tr(T^{a}_{r}T^{b}_{r})=C(r)\delta^{ab}$ and is only independent of $N$ for the fundamental representation. For interacting theories with a slightly broken higher spin symmetry, e.g. the critical $O(N)$ model and $CP^{N-1}$ field theory in 3-dimensions, we also see that $C_J$ stays finite when $N\rightarrow\infty$ \cite{Petkou:1995vu,Huh:2013vga}. In contrast, generalized free theories have $C_J\rightarrow\infty$ and do not have higher spin symmetries. $\mathcal{N}=1$ SQCD in the Veneziano limit has $C_{J}\propto N$ and does not have higher spin symmetries either (except at the boundaries of the conformal window). All these examples are consistent with our sufficient condition for the existence of higher spin currents at infinite $N$.

The relevant crossing equation is the first line of (\ref{eq:iad1}); we reproduce it here:
\begin{equation}
\left(\frac{u}{v}\right)^{\Delta_{\phi}}I_{t}(v,u)=\frac{1}{N}I_{s}(u,v)+\left(1-\frac{1}{N^{2}}\right)Adj_{s}(u,v).
\end{equation}
Since $I_{s}(u,v)$ is at most $\mathcal{O}(1)$, the first term drops
out at large $N$, giving
\begin{equation}
\left(\frac{u}{v}\right)^{\Delta_{\phi}}I_{t}(v,u)=Adj_{s}(u,v)+\mathcal{O}(\frac{1}{N}).\label{eq:LargeNCorssing1}
\end{equation}
We now argue that at $N\rightarrow\infty$, this crossing equation is not solvable in a unitary CFT without an infinite number of conserved higher spin currents. 
We will first consider the case that we do not have scalars with $\Delta\le d-2$ contributing to $Adj_{s}(u,v)$ at large $N$. 
Their effect will be considered in the end of this subsection. 
Without low twist scalars or higher spin currents, the dominant contribution in $Adj_{s}(u,v)$ is given by the conserved global currents: 
\begin{equation}
Adj_{s}(u,v)=\frac{1}{2C_{J}S_{d}^{2}}g_{d-2,1}(u,v)+\dots.
\end{equation}
When $u \ll v \ll1$, we have the behavior
\begin{equation}
g_{d-2,1}(u,v)=\frac{\Gamma(d)}{\Gamma(\frac{d}{2})^{2}}u^{\frac{d-2}{2}}\left(-\log v+2\left(\psi(1)-\psi(\frac{d}{2})\right)\right) + \ldots.
\end{equation}
At small $v$, the $\log v$ term dominates and the crossing equation becomes
\begin{equation}
\left(\frac{u}{v}\right)^{\Delta_{\phi}}I_{t}(v,u)=-\frac{1}{2C_{J}S_{d}^{2}}\frac{\Gamma(d)}{\Gamma(\frac{d}{2})^{2}}u^{\frac{d-2}{2}}\log v+\dots.
\label{eq:LargeNCrossing3}
\end{equation}
It is not possible to reproduce this $\log v$ term on the LHS in a unitary CFT. 
In the reflection positive 4-point function we are considering, each operator should give a positive contribution to $I_t(v,u)$, but matching the $\log v$ term requires negative contributions. We present a detailed argument for this statement in Appendix~\ref{app:LargeNArgument}. 
The only possible remedy for this problem is having an infinite number of higher spin conserved currents contributing to $Adj_{s}(u,v)$.
A finite number of higher spin currents will not solve the problem because they all contribute a $\log v$ term with the same sign.
As discussed in \cite{Alday:2015ota}, the infinite sum over this tower of $\log v$ singularities will yield a power-law singularity in $v$ and the crossing equations can then be solved.\footnote{Another situation where an infinite sum over $\log v$ singularities produces a power law is in the expansion of 2D Virasoro blocks in terms of global conformal blocks. This for example plays an important role in the analysis of 2D Virasoro blocks in~\cite{Fitzpatrick:2014vua,Fitzpatrick:2015qma,Fitzpatrick:2015zha,Fitzpatrick:2015foa}.}

We now consider the contributions of scalar operators with $\Delta_i\le d-2$:
\begin{equation}
Adj_{s}(u,v)={\displaystyle \sum_{i}}P_{i}u^{\frac{\Delta_{i}}{2}}f_{\Delta_{i},0}(0,v)+\frac{1}{2C_{J}S_{d}^{2}}g_{d-2,1}(u,v)+\dots,
\end{equation}
where
\begin{equation}
f_{\Delta_{i},0}(0,v)=\frac{\Gamma(\Delta_{i})}{\Gamma(\frac{\Delta_{i}}{2})^{2}}\left(-\log v+2 \Big(\psi(1)-\psi(\frac{\Delta_{i}}{2})\Big)\right) + \ldots.
\end{equation}
If the smallest $\Delta_{i}$ is less than $d-2$, then we cannot introduce higher spin operators to convert the $\log v$ into power law singularities since they will violate unitarity bounds. 
Therefore the crossing equations cannot be satisfied unless $P_{i}\rightarrow0$ as $N\rightarrow\infty$. 
If $\Delta_{i}=d-2$ with $P_{i}$ finite at large $N$, then we need an infinite number of higher spin currents.

This analysis is not restricted to scalars or conserved operators in the adjoint representation of $SU(N)$. Rather this problem generically arises in CFTs with any large rank global symmetry group when the lowest twist operator whose OPE coefficient is not $\frac{1}{N}$ suppressed does not belong to an infinite tower of states with fixed twist. Similar arguments will also apply e.g. to adjoint and symmetric tensor representations appearing in the OPE of $O(N)$ fundamentals. What makes the case of current contributions special is that $C_{J}$ is a universal quantity which can approach a constant as $N\rightarrow\infty$ in a variety of theories, implying the existence of higher spin currents for these theories. 

In theories where $C_J \sim \cO(1)$ and higher spin currents appear as $N \rightarrow \infty$, it would be very interesting to compute their anomalous dimensions in the $\frac{1}{N}$ expansion using bootstrap techniques. This would require considering situations with slightly broken higher spin symmetry as in~\cite{Maldacena:2012sf, Alday:2015ota}, but taking into account the full global symmetry structure. In particular it would be interesting to understand the role of higher spin currents in different representations of the global symmetry group. We leave a full analysis of this direction to future work.

\section{Discussion}
\label{sec:conclusions}

In this paper, we applied the method of~\cite{Fitzpatrick:2012yx,Komargodski:2012ek} to CFTs with global symmetries in spacetime dimensions $D>2$. We assumed our CFT contains scalar operators transforming under various representations of the global symmetry. Crossing symmetry of their $4$-point functions implies the existence of large spin double-twist operators in all possible symmetry representations. We then computed their anomalous dimensions as a function of the central charge $C_T$, the current central charge $C_J$, and the OPE coefficients of low-dimension scalars. These results correspond in AdS to the binding energies of large spin two-particle states arising from gravitational, gauge, and scalar interactions. 

As expected, we saw that the gravitational binding energy is negative and universal (it does not depend on the representation of the two-particle state). The gauge binding energy, on the other hand, has a sign and magnitude that is determined by the representation. For scalar exchange, the result similarly depends on the representation of the exchanged scalar  -- in particular, when a charged scalar is exchanged, the resulting binding energy for even and odd spin two-particle states can have opposite sign. This is not in conflict with Nachtmann's theorem \cite{Nachtmann:1973mr}, which only implies convexity for the even spin, minimal twist sector in reflection positive OPEs.

We applied our analysis to $4D$ $SU(N)$ $\mathcal{N}=1$ SQCD and the $3D$ $O(N)$ vector models. In both these cases we focused on the limit $N \ll \ell$, where the anomalous dimensions can be used as perturbative parameters. For $\mathcal{N}=1$ SQCD we found the anomalous dimensions for a generic number of flavors and colors within the conformal window and then considered their behavior in the Veneziano limit, where both the number of colors and the number of flavors become large. In this limit we saw that some double-twist anomalous dimensions do not have any additional suppression with $N$ at large spin, which is consistent with the existence of generalized single trace states. For the $O(N)$ vector models we made use of existing numerical results for the conformal dimensions and central charges to write down approximate formulas for the anomalous dimensions.

In addition, we discussed general CFTs with a large global symmetry group and considered the limit $1\ll \ell \ll N$. In this regime we argued that if the current central charge $C_{J}$ does not grow with $N$, then there must exist an infinite tower of higher spin currents in the adjoint representation of the global symmetry group. Such theories also cannot have scalar operators of dimension $\Delta < d-2$ whose OPE coefficients remain finite as $N \rightarrow \infty$. These statements are consistent with known theories that have a higher spin symmetry in the $N\rightarrow\infty$ limit, e.g. the $O(N)$ vector models \cite{Petkou:1995vu}.

In this work our computations were mostly restricted to the regime $N \ll\ell$. To fully extend our analysis to the cases with $\ell \sim N$ and $1\ll \ell \ll N$ with $C_J \sim \cO(1)$, we would need to include the effects of a tower of approximate higher spin currents in the s-channel. This would allow us to more directly make contact with and extend existing results on CFTs with a slightly broken higher spin symmetry~\cite{Maldacena:2012sf,Alday:2015ota}. It would also be interesting to extend our analysis to the Regge limit and to make connections with causality constraints outside the lightcone~\cite{Hartman:2015lfa}. We hope that the present work and its extension to correlators containing currents and stress tensors~\cite{CurrentPaper:2015} will constitute useful steps towards unraveling the beautiful and universal structures inherent to higher-dimensional conformal field theories.

\section*{Acknowledgements}
We thank Tom Hartman, Jared Kaplan, Filip Kos, Juan Maldacena, Jo\~ao Penedones, Balt van Rees, Slava Rychkov, David Simmons-Duffin, Andreas Stergiou, Alessandro Vichi, Junpu Wang, and Sasha Zhiboedov for discussions and Jared Kaplan and David Simmons-Duffin for comments on the draft. This work is supported by NSF grant 1350180.  DP and DL thank the Aspen Center for Physics for its hospitality during the completion of this work, supported by NSF Grant 1066293. DL thanks the Simons Center for Geometry and Physics at Stony Brook University for its hospitality during the completion of this work. DP receives additional support as a Martin A. and Helen Chooljian Founders' Circle Member at the Institute for Advanced Study. DL and DM also thank the Institute for Advanced Study for its hospitality during the completion of this work.

\clearpage
\appendix

\section{Scalar-Scalar OPE}
\label{app:scalarOPE}
In this appendix we review the OPE of two scalars in generic representations of global symmetries. If the two scalars are in conjugate representations, then their OPE takes the form 
\begin{align}
\phi_m(x) \times \phi^n(0) = \frac{\delta_{m}^n}{x^{2\Delta_{\phi}}} -  i(T^a)_{m}^{n} \frac{1}{C_JS_{d}} \frac{x^\mu}{x^{2\Delta_{\phi}-d+2}}  J^{a}_{\mu}(0) + \delta_{m}^n \frac{1}{C_T S_{d}} \frac{\Delta_\phi d}{(d-1)}  \frac{ x^\mu x^\nu}{x^{2\Delta_{\phi}-d+2}}  T_{\mu\nu}(0) + \ldots, 
\end{align}
where $S_d=2 \pi^{d/2}\Gamma(d/2)^{-1}$ is the area of a ($d-1$)-dimensional sphere. The OPE coefficients of the stress tensor and the global symmetry currents are related to the corresponding central charges appearing in their 2-point functions:
\begin{align}
\langle J_{\mu}^{a}(x)J_{\nu}^{b}(0)\rangle &= \delta^{ab}C_{J}\frac{1}{x^{2(d-1)}} I_{\mu\nu},\\
\langle T_{\mu\nu}(x)T_{\rho\sigma}(0)\rangle &= C_{T}\frac{1}{x^{2d}} \mathcal{I}_{\mu\nu,\rho\sigma},
\end{align}
where 
\be
I_{\mu\nu}=\eta_{\mu\nu}-2\frac{x_{\mu}x_{\nu}}{x^{2}},\hspace{1cm}
\mathcal{I}_{\mu\nu,\rho\sigma}=\frac{1}{2}(I_{\mu\rho}I_{\nu\sigma}+I_{\mu\sigma}I_{\nu\rho})-\frac{1}{2}\eta_{\mu\nu}\eta_{\rho\sigma}.
\ee
Our normalization for $J$ and $T$ follow the conventions of \cite{Osborn:1993cr} which does not include a factor of $S_{d}^{-2}$ in the two point function. Note that we have not yet normalized $T$ and $J$ to have 2-point functions $\propto 1$. If we do this, then the OPE coefficients of $\hat{J} = \frac{1}{\sqrt{C_J}} J$ and $\hat{T} = \frac{1}{\sqrt{C_T}} T$ become
\be
(\lambda_{\phi\phi \hat{T}})^n_m=\delta^n_m \frac{d}{d-1}\frac{\Delta_{\phi}}{\sqrt{C_{T}}S_{d}}, \hspace{1.5cm} (\lambda^a_{\phi\phi \hat{J}})^{n}_m=-i(T^a)^n_m \frac{1}{\sqrt{C_{J}}S_{d}}.
\ee
These coefficients are derived from the following Ward identities: 
\begin{align}
\langle \partial^\mu T_{\mu\nu} (x_1) \phi(x_2) \phi(x_3) \rangle &= \delta^d(x_1-x_2)\langle \partial^\nu \phi(x_2) \phi(x_3) \rangle + \delta^d(x_1-x_3)\langle \phi(x_2) \partial^\nu \phi(x_3) \rangle, \\
\langle \partial^\mu J^{a}_{\mu} (x_1) \phi_i(x_2) \phi_j(x_3) \rangle &= \delta^d(x_1-x_2) i(T^a)_{i}^{k} \langle\phi_k(x_2) \phi_j(x_3) \rangle + \delta^d(x_1-x_3)i(T^a)_{j}^{k} \langle \phi_i(x_2) \phi_k(x_3) \rangle.
\end{align}
Our conventions for the non-Abelian generators are as follows. 
The structure constant and the index of a representation is defined as usual:  $[T^a,T^b]=if^{abc}T^c$, $Tr(T^aT^b)=C(r)\delta^{ab}$. 
For each group, we explicitly define the the index of the fundamental representation. 
Matching $f_{abc}=\frac{1}{C(r)} Tr(T^a[T^b,T^c])$ then fixes the normalization of generators for any generic representation. 
The adjoint is given by $i (T_{Adj}^a)^{bc} = f^{abc}$. 
We define the current central charge such that the 3-point function coefficient of canonically normalized conserved currents is equal to $\frac{1}{\sqrt{C_J}S_{d}}f^{abc}$.

For $SU(N)$, we choose the fundamental generator to be  
\be
(T_{i}^{j})_{k}^l\equiv (T^a)_{i}^j (T^a)_{k}^l=\delta_{i}^{l} \delta_{k}^{j} - \frac{1}{N} \delta_{i}^{j} \delta_{k}^{l}.
\ee
Therefore $C=1$. This convention may be slightly different from the standard choice. For example, for the fundamental representation of $SU(2)$, the generators are given by $T^a=\frac{1}{\sqrt{2}}\sigma^a$, the structure constant is $f^{abc}=\sqrt{2}\epsilon^{abc}$. The 4-point projector $T^a T^a$ thus generated matches with what we used in (\ref{eq:SUNFundamentalCrossing}).

We define the generator of the $O(N)$ group on the fundamental representation as 
\be
i(T_{ij})_{kl}\equiv\delta_{ik}\delta_{jl}-\delta_{il}\delta_{jk}. 
\ee
Therefore $C=2$. Note that if interpreted as adjoint indices, the anti-symmetric pair ${ij}$ only runs through three values: ${12}, {13}$, and ${23}$. 
The adjoint generators are obtained by computing the structure constants.
The generators in the symmetric representation are computed by acting with a fundamental generator on each individual index $i (T_{ij})_{kl,mn}= i (T_{ij})_{km} \delta_{ln} + i (T_{ij})_{ln} \delta_{km}$ and then symmetrizing. 

\section{Tensor Structures and Crossing Relations}
\label{app:structs}
\subsection{$O(N)$ Adjoints}

Tensor product representations of the $O(N)$ group break down into irreducible representations (irreps) characterized by traceless tensors with permutation symmetry specified by the Young tableaux. For example, the tensor product of two $O(N)$ adjoints can be decomposed into the following irreps:
\be
r = \bigg( I \,, \tiny\yng(1,1) \,, \yng(2) \,, \yng(2,2) \,, \yng(2,1,1) \,, \yng(1,1,1,1)\,\bigg).
\ee 

The Young projectors can be generated by first anti-symmetrizing indices along the columns and then symmetrizing them along the rows. We also need to eliminate the traces. In this way, we obtain the following generators: 
\be
P_r = \tilde{P}_r - traces,
\ee 
where a Young projector $\tilde{P}_r$ maps an arbitrary 4-index tensor $T_{i_1,i_2,i_3,i_4}$ to a tensor with desired exchange properties. We also 
remove the traces after the projection. More explicitly, $\tilde{P}_r$ are given by:  
\begin{align*}
\tilde{P}_{1}= \ \delta _{i_1 i_3} \delta _{i_2 i_4},
\end{align*}
\begin{align*}
\tilde{P}_{2} =& \ \delta _{i_2 i_4} \delta _{i_3 j_1} \delta _{i_1 j_2}-\delta _{i_2 i_3} \delta _{i_4 j_1} \delta _{i_1 j_2}-\delta _{i_1 i_4} \delta _{i_3
   j_1} \delta _{i_2 j_2}+\delta _{i_1 i_3} \delta _{i_4 j_1} \delta _{i_2 j_2}\\&-\delta _{i_2 i_4} \delta _{i_1 j_1} \delta _{i_3 j_2}+\delta
   _{i_1 i_4} \delta _{i_2 j_1} \delta _{i_3 j_2}+\delta _{i_2 i_3} \delta _{i_1 j_1} \delta _{i_4 j_2}-\delta _{i_1 i_3} \delta _{i_2 j_1}
   \delta _{i_4 j_2},
\end{align*}
\begin{align*}
\tilde{P}_{3} =&-\delta _{i_2 i_4} \delta _{i_3 j_1} \delta _{i_1 j_2}+\delta _{i_2 i_3} \delta _{i_4 j_1} \delta _{i_1 j_2}+\delta _{i_1 i_4} \delta _{i_3
   j_1} \delta _{i_2 j_2}-\delta _{i_1 i_3} \delta _{i_4 j_1} \delta _{i_2 j_2}\\&-\delta _{i_2 i_4} \delta _{i_1 j_1} \delta _{i_3 j_2}+\delta 
   _{i_1 i_4} \delta _{i_2 j_1} \delta _{i_3 j_2}+\delta _{i_2 i_3} \delta _{i_1 j_1} \delta _{i_4 j_2}-\delta _{i_1 i_3} \delta _{i_2 j_1}
   \delta _{i_4 j_2},
\end{align*}
\begin{align*}
\tilde{P}_{4} =& \delta _{i_4 j_1} \delta _{i_3 j_2} \delta _{i_2 j_3} \delta _{i_1 j_4}-\delta _{i_3 j_1} \delta _{i_4 j_2} \delta _{i_2 j_3} \delta _{i_1
   j_4}-\delta _{i_2 j_1} \delta _{i_4 j_2} \delta _{i_3 j_3} \delta _{i_1 j_4}+\delta _{i_2 j_1} \delta _{i_3 j_2} \delta _{i_4 j_3} \delta
   _{i_1 j_4}\\&-\delta _{i_4 j_1} \delta _{i_3 j_2} \delta _{i_1 j_3} \delta _{i_2 j_4}+\delta _{i_3 j_1} \delta _{i_4 j_2} \delta _{i_1 j_3}
   \delta _{i_2 j_4}+\delta _{i_1 j_1} \delta _{i_4 j_2} \delta _{i_3 j_3} \delta _{i_2 j_4}-\delta _{i_1 j_1} \delta _{i_3 j_2} \delta
   _{i_4 j_3} \delta _{i_2 j_4}\\&-\delta _{i_4 j_1} \delta _{i_2 j_2} \delta _{i_1 j_3} \delta _{i_3 j_4}+\delta _{i_4 j_1} \delta _{i_1 j_2}
   \delta _{i_2 j_3} \delta _{i_3 j_4}+\delta _{i_2 j_1} \delta _{i_1 j_2} \delta _{i_4 j_3} \delta _{i_3 j_4}-\delta _{i_1 j_1} \delta
   _{i_2 j_2} \delta _{i_4 j_3} \delta _{i_3 j_4}\\&+\delta _{i_3 j_1} \delta _{i_2 j_2} \delta _{i_1 j_3} \delta _{i_4 j_4}-\delta _{i_3 j_1}
   \delta _{i_1 j_2} \delta _{i_2 j_3} \delta _{i_4 j_4}-\delta _{i_2 j_1} \delta _{i_1 j_2} \delta _{i_3 j_3} \delta _{i_4 j_4}+\delta
   _{i_1 j_1} \delta _{i_2 j_2} \delta _{i_3 j_3} \delta _{i_4 j_4},
\end{align*}
\begin{align*}
\tilde{P}_{5} =&-\delta _{i_4 j_1} \delta _{i_3 j_2} \delta _{i_2 j_3} \delta _{i_1 j_4}+\delta _{i_4 j_1} \delta _{i_2 j_2} \delta _{i_3 j_3} \delta _{i_1
   j_4}+\delta _{i_4 j_1} \delta _{i_3 j_2} \delta _{i_1 j_3} \delta _{i_2 j_4}-\delta _{i_4 j_1} \delta _{i_1 j_2} \delta _{i_3 j_3} \delta
   _{i_2 j_4}\\&-\delta _{i_4 j_1} \delta _{i_2 j_2} \delta _{i_1 j_3} \delta _{i_3 j_4}+\delta _{i_4 j_1} \delta _{i_1 j_2} \delta _{i_2 j_3}
   \delta _{i_3 j_4}-\delta _{i_3 j_1} \delta _{i_2 j_2} \delta _{i_1 j_3} \delta _{i_4 j_4}+\delta _{i_2 j_1} \delta _{i_3 j_2} \delta
   _{i_1 j_3} \delta _{i_4 j_4}\\&+\delta _{i_3 j_1} \delta _{i_1 j_2} \delta _{i_2 j_3} \delta _{i_4 j_4}-\delta _{i_1 j_1} \delta _{i_3 j_2}
   \delta _{i_2 j_3} \delta _{i_4 j_4}-\delta _{i_2 j_1} \delta _{i_1 j_2} \delta _{i_3 j_3} \delta _{i_4 j_4}+\delta _{i_1 j_1} \delta
   _{i_2 j_2} \delta _{i_3 j_3} \delta _{i_4 j_4},
\end{align*}
\begin{align*}
\tilde{P}_{6} =& \delta _{i_4 j_1} \delta _{i_3 j_2} \delta _{i_2 j_3} \delta _{i_1 j_4}-\delta _{i_3 j_1} \delta _{i_4 j_2} \delta _{i_2 j_3} \delta _{i_1
   j_4}-\delta _{i_4 j_1} \delta _{i_2 j_2} \delta _{i_3 j_3} \delta _{i_1 j_4}+\delta _{i_2 j_1} \delta _{i_4 j_2} \delta _{i_3 j_3} \delta
   _{i_1 j_4}\\&+\delta _{i_3 j_1} \delta _{i_2 j_2} \delta _{i_4 j_3} \delta _{i_1 j_4}-\delta _{i_2 j_1} \delta _{i_3 j_2} \delta _{i_4 j_3}
   \delta _{i_1 j_4}-\delta _{i_4 j_1} \delta _{i_3 j_2} \delta _{i_1 j_3} \delta _{i_2 j_4}+\delta _{i_3 j_1} \delta _{i_4 j_2} \delta
   _{i_1 j_3} \delta _{i_2 j_4}\\&+\delta _{i_4 j_1} \delta _{i_1 j_2} \delta _{i_3 j_3} \delta _{i_2 j_4}-\delta _{i_1 j_1} \delta _{i_4 j_2}
   \delta _{i_3 j_3} \delta _{i_2 j_4}-\delta _{i_3 j_1} \delta _{i_1 j_2} \delta _{i_4 j_3} \delta _{i_2 j_4}+\delta _{i_1 j_1} \delta
   _{i_3 j_2} \delta _{i_4 j_3} \delta _{i_2 j_4}\\&+\delta _{i_4 j_1} \delta _{i_2 j_2} \delta _{i_1 j_3} \delta _{i_3 j_4}-\delta _{i_2 j_1}
   \delta _{i_4 j_2} \delta _{i_1 j_3} \delta _{i_3 j_4}-\delta _{i_4 j_1} \delta _{i_1 j_2} \delta _{i_2 j_3} \delta _{i_3 j_4}+\delta
   _{i_1 j_1} \delta _{i_4 j_2} \delta _{i_2 j_3} \delta _{i_3 j_4}\\&+\delta _{i_2 j_1} \delta _{i_1 j_2} \delta _{i_4 j_3} \delta _{i_3
   j_4}-\delta _{i_1 j_1} \delta _{i_2 j_2} \delta _{i_4 j_3} \delta _{i_3 j_4}-\delta _{i_3 j_1} \delta _{i_2 j_2} \delta _{i_1 j_3} \delta
   _{i_4 j_4}+\delta _{i_2 j_1} \delta _{i_3 j_2} \delta _{i_1 j_3} \delta _{i_4 j_4}\\&+\delta _{i_3 j_1} \delta _{i_1 j_2} \delta _{i_2 j_3}
   \delta _{i_4 j_4}-\delta _{i_1 j_1} \delta _{i_3 j_2} \delta _{i_2 j_3} \delta _{i_4 j_4}-\delta _{i_2 j_1} \delta _{i_1 j_2} \delta
   _{i_3 j_3} \delta _{i_4 j_4}+\delta _{i_1 j_1} \delta _{i_2 j_2} \delta _{i_3 j_3} \delta _{i_4 j_4}.
\end{align*}
The three point structures are given by, for example:
\be
&& \langle A_{i_1 i_2} A_{i_3,i_4}  (O_r)_{j_1,j_2,j_3,j_4} \rangle   \propto  S_{i_1 i_2 ; i_3 i4; j_1,j_2,j_3,j_4}, \nn\\
&& S^{A,r}_{i_1 i_2 ; i_3 i4; j_1, j_2,j_3,j_4} = 
\frac{1}{4}(\delta_{i_1 i^{\prime}_1}\delta_{i_2 i^{\prime}_2}-\delta_{i_1 i^{\prime}_2}\delta_{i_2 i^{\prime}_1})
(\delta_{i_3 i^{\prime}_3}\delta_{i_4 i^{\prime}_4}-\delta_{i_3 i^{\prime}_4}\delta_{i_4 i^{\prime}_3})(P_r)_{i^{\prime}_1 i^{\prime}_2 i^{\prime}_3 i^{\prime}_4;j_1 j_2 j_3 j_4}.
\ee
We then build the 4-point structures $t^{A,r}$ in $\<A_{i_1 i_2}A_{i_3 i_4} A_{j_1 j_2} A_{j_3 j_4}\>$ by contracting two 3-point structures together: 
\be
t^{A,r}_{i_1 i_2 ; i_3 i4; j_1 j_2 ; j_3 j_4} =\frac{1}{n_r} S^{A,r}_{i_1 i_2 ; i_3 i_4; j^{\prime}_1 j^{\prime}_2 j^{\prime}_3 j^{\prime}_4} S^{A,r}_{j_1 j_2 ; j_3 j_4; j^{\prime}_1 j^{\prime}_2 j^{\prime}_3 j^{\prime}_4},
\ee
where $n_r=(1/4,-2,2,24,-6,16)$ are normalizations chosen for the 4-point structures. They are chosen such that certain reflection positive configurations have unit tensor structure. In particular, $t^{A,1}_{1212;1212} = 1$,  $t^{A,r}_{1213;1312} = 1$ for $r=2,3$, and  $t^{A,r}_{1234;3412} = 1$ for $r=4,5,6$. We explicitly write down $t^{A,r}$ for $r=1,2,3$ below. Note that we permuted the indices such that the 4-point function is $\<A_{i_1 j_1}A_{i_2 j_2} A_{i_3 j_3} A_{i_4 j_4}\>$.
\begin{align*}
t^{A,1} =& \left(\delta _{i_2 i_3} \delta _{i_1 i_4}-\delta _{i_1 i_3} \delta _{i_2 i_4}\right) \left(\delta _{j_2 j_3} \delta _{j_1 j_4}-\delta _{j_1 j_3} \delta _{j_2
   j_4}\right),
\end{align*}
\begin{align*}
t^{A,2} =& -\delta _{i_2 i_4} \delta _{i_3 j_2} \delta _{j_1 j_3} \delta _{i_1 j_4}-\delta _{i_2 i_3} \delta _{i_4 j_2} \delta _{j_1 j_3} \delta _{i_1 j_4}-\delta _{i_2 i_4}
   \delta _{i_3 j_1} \delta _{j_2 j_3} \delta _{i_1 j_4}+\delta _{i_2 i_3} \delta _{i_4 j_1} \delta _{j_2 j_3} \delta _{i_1 j_4}\\&-\delta _{i_1 i_4} \delta _{i_3
   j_2} \delta _{j_1 j_3} \delta _{i_2 j_4}+\delta _{i_1 i_3} \delta _{i_4 j_2} \delta _{j_1 j_3} \delta _{i_2 j_4}+\delta _{i_1 i_4} \delta _{i_3 j_1} \delta
   _{j_2 j_3} \delta _{i_2 j_4}-\delta _{i_1 i_3} \delta _{i_4 j_1} \delta _{j_2 j_3} \delta _{i_2 j_4}\\&-\delta _{i_2 i_4} \delta _{i_1 j_2} \delta _{j_1 j_3}
   \delta _{i_3 j_4}+\delta _{i_1 i_4} \delta _{i_2 j_2} \delta _{j_1 j_3} \delta _{i_3 j_4}+\delta _{i_2 i_4} \delta _{i_1 j_1} \delta _{j_2 j_3} \delta _{i_3
   j_4}-\delta _{i_1 i_4} \delta _{i_2 j_1} \delta _{j_2 j_3} \delta _{i_3 j_4}\\&+\delta _{i_2 i_3} \delta _{i_1 j_2} \delta _{j_1 j_3} \delta _{i_4 j_4}-\delta
   _{i_1 i_3} \delta _{i_2 j_2} \delta _{j_1 j_3} \delta _{i_4 j_4}-\delta _{i_2 i_3} \delta _{i_1 j_1} \delta _{j_2 j_3} \delta _{i_4 j_4}+\delta _{i_1 i_3}
   \delta _{i_2 j_1} \delta _{j_2 j_3} \delta _{i_4 j_4}\\&-\delta _{i_2 i_4} \delta _{i_3 j_2} \delta _{i_1 j_3} \delta _{j_1 j_4}+\delta _{i_2 i_3} \delta _{i_4
   j_2} \delta _{i_1 j_3} \delta _{j_1 j_4}+\delta _{i_1 i_4} \delta _{i_3 j_2} \delta _{i_2 j_3} \delta _{j_1 j_4}-\delta _{i_1 i_3} \delta _{i_4 j_2} \delta
   _{i_2 j_3} \delta _{j_1 j_4}\\&+\delta _{i_2 i_4} \delta _{i_1 j_2} \delta _{i_3 j_3} \delta _{j_1 j_4}-\delta _{i_1 i_4} \delta _{i_2 j_2} \delta _{i_3 j_3}
   \delta _{j_1 j_4}-\delta _{i_2 i_3} \delta _{i_1 j_2} \delta _{i_4 j_3} \delta _{j_1 j_4}+\delta _{i_1 i_3} \delta _{i_2 j_2} \delta _{i_4 j_3} \delta _{j_1
   j_4}\\&+\delta _{i_2 i_4} \delta _{i_3 j_1} \delta _{i_1 j_3} \delta _{j_2 j_4}-\delta _{i_2 i_3} \delta _{i_4 j_1} \delta _{i_1 j_3} \delta _{j_2 j_4}-\delta
   _{i_1 i_4} \delta _{i_3 j_1} \delta _{i_2 j_3} \delta _{j_2 j_4}+\delta _{i_1 i_3} \delta _{i_4 j_1} \delta _{i_2 j_3} \delta _{j_2 j_4}\\&-\delta _{i_2 i_4}
   \delta _{i_1 j_1} \delta _{i_3 j_3} \delta _{j_2 j_4}+\delta _{i_1 i_4} \delta _{i_2 j_1} \delta _{i_3 j_3} \delta _{j_2 j_4}+\delta _{i_2 i_3} \delta _{i_1
   j_1} \delta _{i_4 j_3} \delta _{j_2 j_4}-\delta _{i_1 i_3} \delta _{i_2 j_1} \delta _{i_4 j_3} \delta _{j_2 j_4},
\end{align*}
\begin{align*}
t^{A,3} =& \delta _{i_2 i_4} \delta _{i_3 j_2} \delta _{j_1 j_3} \delta _{i_1 j_4}-\delta _{i_2 i_3} \delta _{i_4 j_2} \delta _{j_1 j_3} \delta _{i_1 j_4}-\delta _{i_2 i_4}
   \delta _{i_3 j_1} \delta _{j_2 j_3} \delta _{i_1 j_4}+\delta _{i_2 i_3} \delta _{i_4 j_1} \delta _{j_2 j_3} \delta _{i_1 j_4}\\&
   -\delta _{i_1 i_4} \delta _{i_3 j_2} \delta _{j_1 j_3} \delta _{i_2 j_4}+\delta _{i_1 i_3} \delta _{i_4 j_2} \delta _{j_1 j_3} \delta _{i_2 j_4}+\delta _{i_1 i_4} \delta _{i_3 j_1} \delta
   _{j_2 j_3} \delta _{i_2 j_4}-\delta _{i_1 i_3} \delta _{i_4 j_1} \delta _{j_2 j_3} \delta _{i_2 j_4}\\&+\delta _{i_2 i_4} \delta _{i_1 j_2} \delta _{j_1 j_3}
   \delta _{i_3 j_4}-\delta _{i_1 i_4} \delta _{i_2 j_2} \delta _{j_1 j_3} \delta _{i_3 j_4}-\delta _{i_2 i_4} \delta _{i_1 j_1} \delta _{j_2 j_3} \delta _{i_3
   j_4}+\delta _{i_1 i_4} \delta _{i_2 j_1} \delta _{j_2 j_3} \delta _{i_3 j_4}\\&-\delta _{i_2 i_3} \delta _{i_1 j_2} \delta _{j_1 j_3} \delta _{i_4 j_4}+\delta
   _{i_1 i_3} \delta _{i_2 j_2} \delta _{j_1 j_3} \delta _{i_4 j_4}+\delta _{i_2 i_3} \delta _{i_1 j_1} \delta _{j_2 j_3} \delta _{i_4 j_4}-\delta _{i_1 i_3}
   \delta _{i_2 j_1} \delta _{j_2 j_3} \delta _{i_4 j_4}\\&-\delta _{i_2 i_4} \delta _{i_3 j_2} \delta _{i_1 j_3} \delta _{j_1 j_4}+\delta _{i_2 i_3} \delta _{i_4
   j_2} \delta _{i_1 j_3} \delta _{j_1 j_4}+\delta _{i_1 i_4} \delta _{i_3 j_2} \delta _{i_2 j_3} \delta _{j_1 j_4}-\delta _{i_1 i_3} \delta _{i_4 j_2} \delta
   _{i_2 j_3} \delta _{j_1 j_4}\\&-\delta _{i_2 i_4} \delta _{i_1 j_2} \delta _{i_3 j_3} \delta _{j_1 j_4}+\delta _{i_1 i_4} \delta _{i_2 j_2} \delta _{i_3 j_3}
   \delta _{j_1 j_4}+\delta _{i_2 i_3} \delta _{i_1 j_2} \delta _{i_4 j_3} \delta _{j_1 j_4}-\delta _{i_1 i_3} \delta _{i_2 j_2} \delta _{i_4 j_3} \delta _{j_1
   j_4}\\&+\delta _{i_2 i_4} \delta _{i_3 j_1} \delta _{i_1 j_3} \delta _{j_2 j_4}-\delta _{i_2 i_3} \delta _{i_4 j_1} \delta _{i_1 j_3} \delta _{j_2 j_4}-\delta
   _{i_1 i_4} \delta _{i_3 j_1} \delta _{i_2 j_3} \delta _{j_2 j_4}+\delta _{i_1 i_3} \delta _{i_4 j_1} \delta _{i_2 j_3} \delta _{j_2 j_4}\\&+\delta _{i_2 i_4}
   \delta _{i_1 j_1} \delta _{i_3 j_3} \delta _{j_2 j_4}-\delta _{i_1 i_4} \delta _{i_2 j_1} \delta _{i_3 j_3} \delta _{j_2 j_4}-\delta _{i_2 i_3} \delta _{i_1
   j_1} \delta _{i_4 j_3} \delta _{j_2 j_4}+\delta _{i_1 i_3} \delta _{i_2 j_1} \delta _{i_4 j_3} \delta _{j_2 j_4}\\&+\frac{8}{N}( \delta _{i_1 i_3} \delta _{i_2 i_4} \delta _{j_2 j_3} \delta _{j_1 j_4}- \delta _{i_2 i_3} \delta _{i_1 i_4} \delta _{j_2 j_3} \delta _{j_1 j_4}+ \delta _{i_2 i_3} \delta _{i_1 i_4} \delta _{j_1
   j_3} \delta _{j_2 j_4}- \delta _{i_1 i_3} \delta _{i_2 i_4} \delta _{j_1 j_3} \delta _{j_2 j_4}).
\end{align*}
As noted in Appendix~\ref{app:scalarOPE}, the generators are normalized as $(T_{ij})_{kl}=\delta_{ik}\delta_{jl}-\delta_{il}\delta_{kj}$. We then have $f_{i_1j_1i_2j_2i_3j_3}=-\frac{i}{2}\Tr\{T_{i_1j_1},[T_{i_2j_2},T_{i_3j_3}]\}$ and $i(T^{Adj}_{i_1j_1})_{i_2j_2i_3j_3}=f_{i_1j_1i_2j_2i_3j_3}$. The contracted generators $\frac{i}{2}(T^{Adj}_{ij})i(T^{Adj}_{ji})$ match with the projector $t^{A,2}$ given above.

Matching tensor structures between the $(12)-(34)$ channel and the $(14)-(32)$ channel gives the crossing relations:
\be
\left(\frac{u}{v}\right)^{\Delta_{\Phi}} G_{r,t}(v,u) = \cM_r^{r'} G_{r',s}(u,v),
\ee
with
\be
\cM^{r'}_{r} &=& \left(
\begin{array}{cccccc}
 \frac{2}{(N-1) N} & \frac{4 (N-2)}{(N-1) N} & \frac{4 (N-2) (N+2)}{(N-1) N^2} & \frac{(N-3) (N+1) (N+2)}{(N-1)^2 N} & \frac{(N-3) (N+2)}{(N-1) N} & \frac{(N-3)
   (N-2)}{(N-1) N} \\
 \frac{1}{2 (N-2)} & \frac{1}{2} & \frac{(N-4) (N+2)}{2 (N-2) N} & -\frac{(N-3) (N+1) (N+2)}{4 (N-2)^2 (N-1)} & 0 & \frac{N-3}{2 (N-2)} \\
 \frac{1}{2 (N-2)} & \frac{N-4}{2 (N-2)} & \frac{N^2-8}{2 (N-2) N} & \frac{(N-4) (N-3) (N+1)}{4 (N-2)^2 (N-1)} & -\frac{N-3}{(N-2)^2} & -\frac{N-3}{2 (N-2)} \\
 \frac{1}{3} & -\frac{2}{3} & \frac{2 (N-4)}{3 N} & \frac{N^2-6 N+11}{3 (N-2) (N-1)} & -\frac{N-4}{3 (N-2)} & \frac{1}{3} \\
 \frac{1}{2} & 0 & -\frac{4}{N} & -\frac{(N-4) (N+1)}{2 (N-2) (N-1)} & \frac{1}{2} & -\frac{1}{2} \\
 \frac{1}{6} & \frac{2}{3} & -\frac{2 (N+2)}{3 N} & \frac{(N+1) (N+2)}{6 (N-2) (N-1)} & -\frac{N+2}{6 (N-2)} & \frac{1}{6} \\
\end{array}
\right).\nn\\
\ee

\subsection{$O(N)$ Symmetric Tensors}
The tensor product of two $O(N)$ symmetric traceless tensors can be decomposed in the following irreps:
\be
r = \bigg( I \,, \tiny\yng(1,1) \,, \yng(2) \,, \yng(2,2) \,, \yng(3,1) \,, \yng(4)\,\bigg).
\ee 
Their Young projectors are: 

\begin{align*}
\tilde{P}_{S,1}= \ \delta _{i_1 i_3} \delta _{i_2 i_4},
\end{align*}
\begin{align*}
\tilde{P}_{S,2} =&-\delta _{i_3 i_4} \delta _{i_2 j_1} \delta _{i_1 j_2}-\delta _{i_2 i_4} \delta _{i_3 j_1} \delta _{i_1 j_2}-2 \delta _{i_2 i_3} \delta
   _{i_4 j_1} \delta _{i_1 j_2}+\delta _{i_3 i_4} \delta _{i_1 j_1} \delta _{i_2 j_2}-\delta _{i_1 i_3} \delta _{i_4 j_1} \delta _{i_2
   j_2}\\&+\delta _{i_2 i_4} \delta _{i_1 j_1} \delta _{i_3 j_2}-\delta _{i_1 i_2} \delta _{i_4 j_1} \delta _{i_3 j_2}+2 \delta _{i_2 i_3}
   \delta _{i_1 j_1} \delta _{i_4 j_2}+\delta _{i_1 i_3} \delta _{i_2 j_1} \delta _{i_4 j_2}+\delta _{i_1 i_2} \delta _{i_3 j_1} \delta
   _{i_4 j_2},
\end{align*}
\begin{align*}
\tilde{P}_{S,3} =& \ \delta_{i_3 i_4} \delta _{i_2 j_1} \delta _{i_1 j_2}+\delta _{i_2 i_4} \delta _{i_3 j_1} \delta _{i_1 j_2}+\delta _{i_2 i_3} \delta _{i_4
   j_1} \delta _{i_1 j_2}+\delta _{i_3 i_4} \delta _{i_1 j_1} \delta _{i_2 j_2}+\delta _{i_1 i_4} \delta _{i_3 j_1} \delta _{i_2 j_2}+\delta
   _{i_1 i_3} \delta _{i_4 j_1} \delta _{i_2 j_2}\\&+\delta _{i_2 i_4} \delta _{i_1 j_1} \delta _{i_3 j_2}+\delta _{i_1 i_4} \delta _{i_2 j_1}
   \delta _{i_3 j_2}+\delta _{i_1 i_2} \delta _{i_4 j_1} \delta _{i_3 j_2}+\delta _{i_2 i_3} \delta _{i_1 j_1} \delta _{i_4 j_2}+\delta
   _{i_1 i_3} \delta _{i_2 j_1} \delta _{i_4 j_2}+\delta _{i_1 i_2} \delta _{i_3 j_1} \delta _{i_4 j_2},
\end{align*}
\begin{align*}
\tilde{P}_{S,4}=& \ \delta _{i_4 j_1} \delta _{i_3 j_2} \delta _{i_2 j_3} \delta _{i_1 j_4}+\delta _{i_3 j_1} \delta _{i_4 j_2} \delta _{i_2 j_3} \delta _{i_1
   j_4}-\delta _{i_3 j_1} \delta _{i_2 j_2} \delta _{i_4 j_3} \delta _{i_1 j_4}-\delta _{i_2 j_1} \delta _{i_3 j_2} \delta _{i_4 j_3} \delta
   _{i_1 j_4}\\&+\delta _{i_4 j_1} \delta _{i_3 j_2} \delta _{i_1 j_3} \delta _{i_2 j_4}+\delta _{i_3 j_1} \delta _{i_4 j_2} \delta _{i_1 j_3}
   \delta _{i_2 j_4}-\delta _{i_4 j_1} \delta _{i_1 j_2} \delta _{i_3 j_3} \delta _{i_2 j_4}-\delta _{i_1 j_1} \delta _{i_4 j_2} \delta
   _{i_3 j_3} \delta _{i_2 j_4}\\&-\delta _{i_4 j_1} \delta _{i_1 j_2} \delta _{i_2 j_3} \delta _{i_3 j_4}-\delta _{i_1 j_1} \delta _{i_4 j_2}
   \delta _{i_2 j_3} \delta _{i_3 j_4}+\delta _{i_2 j_1} \delta _{i_1 j_2} \delta _{i_4 j_3} \delta _{i_3 j_4}+\delta _{i_1 j_1} \delta
   _{i_2 j_2} \delta _{i_4 j_3} \delta _{i_3 j_4}\\&-\delta _{i_3 j_1} \delta _{i_2 j_2} \delta _{i_1 j_3} \delta _{i_4 j_4}-\delta _{i_2 j_1}
   \delta _{i_3 j_2} \delta _{i_1 j_3} \delta _{i_4 j_4}+\delta _{i_2 j_1} \delta _{i_1 j_2} \delta _{i_3 j_3} \delta _{i_4 j_4}+\delta
   _{i_1 j_1} \delta _{i_2 j_2} \delta _{i_3 j_3} \delta _{i_4 j_4},
\end{align*}
\begin{align*}
\tilde{P}_{S,5} =&-\delta _{i_4 j_1} \delta _{i_3 j_2} \delta _{i_2 j_3} \delta _{i_1 j_4}-\delta _{i_3 j_1} \delta _{i_4 j_2} \delta _{i_2 j_3} \delta _{i_1
   j_4}-\delta _{i_4 j_1} \delta _{i_2 j_2} \delta _{i_3 j_3} \delta _{i_1 j_4}-\delta _{i_2 j_1} \delta _{i_4 j_2} \delta _{i_3 j_3} \delta
   _{i_1 j_4}\\&-\delta _{i_3 j_1} \delta _{i_2 j_2} \delta _{i_4 j_3} \delta _{i_1 j_4}-\delta _{i_2 j_1} \delta _{i_3 j_2} \delta _{i_4 j_3}
   \delta _{i_1 j_4}+\delta _{i_3 j_1} \delta _{i_2 j_2} \delta _{i_1 j_3} \delta _{i_4 j_4}+\delta _{i_2 j_1} \delta _{i_3 j_2} \delta
   _{i_1 j_3} \delta _{i_4 j_4}\\&+\delta _{i_3 j_1} \delta _{i_1 j_2} \delta _{i_2 j_3} \delta _{i_4 j_4}+\delta _{i_1 j_1} \delta _{i_3 j_2}
   \delta _{i_2 j_3} \delta _{i_4 j_4}+\delta _{i_2 j_1} \delta _{i_1 j_2} \delta _{i_3 j_3} \delta _{i_4 j_4}+\delta _{i_1 j_1} \delta
   _{i_2 j_2} \delta _{i_3 j_3} \delta _{i_4 j_4},
\end{align*}
\begin{align*}
\tilde{P}_{S,6} =& \ \delta _{i_4 j_1} \delta _{i_3 j_2} \delta _{i_2 j_3} \delta _{i_1 j_4}+\delta _{i_3 j_1} \delta _{i_4 j_2} \delta _{i_2 j_3} \delta _{i_1
   j_4}+\delta _{i_4 j_1} \delta _{i_2 j_2} \delta _{i_3 j_3} \delta _{i_1 j_4}+\delta _{i_2 j_1} \delta _{i_4 j_2} \delta _{i_3 j_3} \delta
   _{i_1 j_4}\\&+\delta _{i_3 j_1} \delta _{i_2 j_2} \delta _{i_4 j_3} \delta _{i_1 j_4}+\delta _{i_2 j_1} \delta _{i_3 j_2} \delta _{i_4 j_3}
   \delta _{i_1 j_4}+\delta _{i_4 j_1} \delta _{i_3 j_2} \delta _{i_1 j_3} \delta _{i_2 j_4}+\delta _{i_3 j_1} \delta _{i_4 j_2} \delta
   _{i_1 j_3} \delta _{i_2 j_4}\\&+\delta _{i_4 j_1} \delta _{i_1 j_2} \delta _{i_3 j_3} \delta _{i_2 j_4}+\delta _{i_1 j_1} \delta _{i_4 j_2}
   \delta _{i_3 j_3} \delta _{i_2 j_4}+\delta _{i_3 j_1} \delta _{i_1 j_2} \delta _{i_4 j_3} \delta _{i_2 j_4}+\delta _{i_1 j_1} \delta
   _{i_3 j_2} \delta _{i_4 j_3} \delta _{i_2 j_4}\\&+\delta _{i_4 j_1} \delta _{i_2 j_2} \delta _{i_1 j_3} \delta _{i_3 j_4}+\delta _{i_2 j_1}
   \delta _{i_4 j_2} \delta _{i_1 j_3} \delta _{i_3 j_4}+\delta _{i_4 j_1} \delta _{i_1 j_2} \delta _{i_2 j_3} \delta _{i_3 j_4}+\delta
   _{i_1 j_1} \delta _{i_4 j_2} \delta _{i_2 j_3} \delta _{i_3 j_4}\\&+\delta _{i_2 j_1} \delta _{i_1 j_2} \delta _{i_4 j_3} \delta _{i_3
   j_4}+\delta _{i_1 j_1} \delta _{i_2 j_2} \delta _{i_4 j_3} \delta _{i_3 j_4}+\delta _{i_3 j_1} \delta _{i_2 j_2} \delta _{i_1 j_3} \delta
   _{i_4 j_4}+\delta _{i_2 j_1} \delta _{i_3 j_2} \delta _{i_1 j_3} \delta _{i_4 j_4}\\&+\delta _{i_3 j_1} \delta _{i_1 j_2} \delta _{i_2 j_3}
   \delta _{i_4 j_4}+\delta _{i_1 j_1} \delta _{i_3 j_2} \delta _{i_2 j_3} \delta _{i_4 j_4}+\delta _{i_2 j_1} \delta _{i_1 j_2} \delta
   _{i_3 j_3} \delta _{i_4 j_4}+\delta _{i_1 j_1} \delta _{i_2 j_2} \delta _{i_3 j_3} \delta _{i_4 j_4}.
\end{align*}

We proceed similarly to the anti-symmetric case. The 3-point structures are constructed by removing the traces from the Young projectors and contracting with anti-symmetric projectors $\frac{1}{2}(\delta_{i_1 i^{\prime}_1}\delta_{i_2 i^{\prime}_2}+\delta_{i_1 i^{\prime}_2}\delta_{i_2 i^{\prime}_1}-\frac{2}{N}\delta_{i_1 i_2}\delta_{i^{\prime}_1 i^{\prime}_2})$. The 4-point structures can be obtained by contracting two 3-point function structures with normalization constants $n_{S,r}=(1/4,-2,2,16,-4,24)$, chosen such that $t^{S,1}_{1212;1212} = 1$,  $t^{S,r}_{1213;1312} = 1$ for $r=2,3$, and $t^{S,r}_{1234;3412} = 1$ for $r=4,5,6$. We explicitly write down $t^{S,r}$ for $r=1,2,3$ below. Note that we slightly permuted the indices such that the 4-point function is $\<S_{i_1 j_1}S_{i_2 j_2} S_{i_3 j_3} S_{i_4 j_4}\>$.

\begin{align*}
t^{S,1} =& \left(\delta _{i_2 i_3} \delta _{i_1 i_4}+\delta _{i_1 i_3} \delta _{i_2 i_4}\right) \left(\delta _{j_2 j_3} \delta _{j_1 j_4}+\delta _{j_1 j_3} \delta _{j_2
   j_4}\right)+\frac{4}{N^2} \left( \delta _{i_1 i_2} \delta _{i_3 i_4} \delta _{j_1 j_2} \delta _{j_3 j_4}\right)\\&-\frac{2}{N} \left(\left(\delta _{i_1 i_2} \delta _{i_3 i_4} \left(\delta
   _{j_2 j_3} \delta _{j_1 j_4}+\delta _{j_1 j_3} \delta _{j_2 j_4}\right)+\left(\delta _{i_2 i_3} \delta _{i_1 i_4}+\delta _{i_1 i_3} \delta _{i_2 i_4}\right)
   \delta _{j_1 j_2} \delta _{j_3 j_4}\right)\right),
\end{align*}
\begin{align*}
t^{S,2} =& \delta _{i_2 i_4} \delta _{i_3 j_2} \delta _{j_1 j_3} \delta _{i_1 j_4}+\delta _{i_2 i_3} \delta _{i_4 j_2} \delta _{j_1 j_3} \delta _{i_1 j_4}+\delta _{i_2 i_4}
   \delta _{i_3 j_1} \delta _{j_2 j_3} \delta _{i_1 j_4}+\delta _{i_2 i_3} \delta _{i_4 j_1} \delta _{j_2 j_3} \delta _{i_1 j_4}\\&+\delta _{i_1 i_4} \delta _{i_3
   j_2} \delta _{j_1 j_3} \delta _{i_2 j_4}+\delta _{i_1 i_3} \delta _{i_4 j_2} \delta _{j_1 j_3} \delta _{i_2 j_4}+\delta _{i_1 i_4} \delta _{i_3 j_1} \delta
   _{j_2 j_3} \delta _{i_2 j_4}+\delta _{i_1 i_3} \delta _{i_4 j_1} \delta _{j_2 j_3} \delta _{i_2 j_4}\\&-\delta _{i_2 i_4} \delta _{i_1 j_2} \delta _{j_1 j_3}
   \delta _{i_3 j_4}-\delta _{i_1 i_4} \delta _{i_2 j_2} \delta _{j_1 j_3} \delta _{i_3 j_4}-\delta _{i_2 i_4} \delta _{i_1 j_1} \delta _{j_2 j_3} \delta _{i_3
   j_4}-\delta _{i_1 i_4} \delta _{i_2 j_1} \delta _{j_2 j_3} \delta _{i_3 j_4}\\&-\delta _{i_2 i_3} \delta _{i_1 j_2} \delta _{j_1 j_3} \delta _{i_4 j_4}-\delta
   _{i_1 i_3} \delta _{i_2 j_2} \delta _{j_1 j_3} \delta _{i_4 j_4}-\delta _{i_2 i_3} \delta _{i_1 j_1} \delta _{j_2 j_3} \delta _{i_4 j_4}-\delta _{i_1 i_3}
   \delta _{i_2 j_1} \delta _{j_2 j_3} \delta _{i_4 j_4}\\&+\delta _{i_2 i_4} \delta _{i_3 j_2} \delta _{i_1 j_3} \delta _{j_1 j_4}+\delta _{i_2 i_3} \delta _{i_4
   j_2} \delta _{i_1 j_3} \delta _{j_1 j_4}+\delta _{i_1 i_4} \delta _{i_3 j_2} \delta _{i_2 j_3} \delta _{j_1 j_4}+\delta _{i_1 i_3} \delta _{i_4 j_2} \delta
   _{i_2 j_3} \delta _{j_1 j_4}\\&-\delta _{i_2 i_4} \delta _{i_1 j_2} \delta _{i_3 j_3} \delta _{j_1 j_4}-\delta _{i_1 i_4} \delta _{i_2 j_2} \delta _{i_3 j_3}
   \delta _{j_1 j_4}-\delta _{i_2 i_3} \delta _{i_1 j_2} \delta _{i_4 j_3} \delta _{j_1 j_4}-\delta _{i_1 i_3} \delta _{i_2 j_2} \delta _{i_4 j_3} \delta _{j_1
   j_4}\\&+\delta _{i_2 i_4} \delta _{i_3 j_1} \delta _{i_1 j_3} \delta _{j_2 j_4}+\delta _{i_2 i_3} \delta _{i_4 j_1} \delta _{i_1 j_3} \delta _{j_2 j_4}+\delta
   _{i_1 i_4} \delta _{i_3 j_1} \delta _{i_2 j_3} \delta _{j_2 j_4}+\delta _{i_1 i_3} \delta _{i_4 j_1} \delta _{i_2 j_3} \delta _{j_2 j_4}\\&-\delta _{i_2 i_4}
   \delta _{i_1 j_1} \delta _{i_3 j_3} \delta _{j_2 j_4}-\delta _{i_1 i_4} \delta _{i_2 j_1} \delta _{i_3 j_3} \delta _{j_2 j_4}-\delta _{i_2 i_3} \delta _{i_1
   j_1} \delta _{i_4 j_3} \delta _{j_2 j_4}-\delta _{i_1 i_3} \delta _{i_2 j_1} \delta _{i_4 j_3} \delta _{j_2 j_4},
\end{align*}
\begin{align*}
t^{S,3} =& \delta _{i_2 i_4} \delta _{i_3 j_2} \delta _{j_1 j_3} \delta _{i_1 j_4}+\delta _{i_2 i_3} \delta _{i_4 j_2} \delta _{j_1 j_3} \delta _{i_1 j_4}+\delta _{i_2 i_4}
   \delta _{i_3 j_1} \delta _{j_2 j_3} \delta _{i_1 j_4}+\delta _{i_2 i_3} \delta _{i_4 j_1} \delta _{j_2 j_3} \delta _{i_1 j_4}\\&+\delta _{i_1 i_4} \delta _{i_3
   j_2} \delta _{j_1 j_3} \delta _{i_2 j_4}+\delta _{i_1 i_3} \delta _{i_4 j_2} \delta _{j_1 j_3} \delta _{i_2 j_4}+\delta _{i_1 i_4} \delta _{i_3 j_1} \delta
   _{j_2 j_3} \delta _{i_2 j_4}+\delta _{i_1 i_3} \delta _{i_4 j_1} \delta _{j_2 j_3} \delta _{i_2 j_4}\\&+\delta _{i_2 i_4} \delta _{i_1 j_2} \delta _{j_1 j_3}
   \delta _{i_3 j_4}+\delta _{i_1 i_4} \delta _{i_2 j_2} \delta _{j_1 j_3} \delta _{i_3 j_4}+\delta _{i_2 i_4} \delta _{i_1 j_1} \delta _{j_2 j_3} \delta _{i_3
   j_4}+\delta _{i_1 i_4} \delta _{i_2 j_1} \delta _{j_2 j_3} \delta _{i_3 j_4}\\&+\delta _{i_2 i_3} \delta _{i_1 j_2} \delta _{j_1 j_3} \delta _{i_4 j_4}+\delta
   _{i_1 i_3} \delta _{i_2 j_2} \delta _{j_1 j_3} \delta _{i_4 j_4}+\delta _{i_2 i_3} \delta _{i_1 j_1} \delta _{j_2 j_3} \delta _{i_4 j_4}+\delta _{i_1 i_3}
   \delta _{i_2 j_1} \delta _{j_2 j_3} \delta _{i_4 j_4}\\&+\delta _{i_2 i_4} \delta _{i_3 j_2} \delta _{i_1 j_3} \delta _{j_1 j_4}+\delta _{i_2 i_3} \delta _{i_4
   j_2} \delta _{i_1 j_3} \delta _{j_1 j_4}+\delta _{i_1 i_4} \delta _{i_3 j_2} \delta _{i_2 j_3} \delta _{j_1 j_4}+\delta _{i_1 i_3} \delta _{i_4 j_2} \delta
   _{i_2 j_3} \delta _{j_1 j_4}\\&+\delta _{i_2 i_4} \delta _{i_1 j_2} \delta _{i_3 j_3} \delta _{j_1 j_4}+\delta _{i_1 i_4} \delta _{i_2 j_2} \delta _{i_3 j_3}
   \delta _{j_1 j_4}+\delta _{i_2 i_3} \delta _{i_1 j_2} \delta _{i_4 j_3} \delta _{j_1 j_4}+\delta _{i_1 i_3} \delta _{i_2 j_2} \delta _{i_4 j_3} \delta _{j_1
   j_4}\\&+\delta _{i_2 i_4} \delta _{i_3 j_1} \delta _{i_1 j_3} \delta _{j_2 j_4}+\delta _{i_2 i_3} \delta _{i_4 j_1} \delta _{i_1 j_3} \delta _{j_2 j_4}+\delta
   _{i_1 i_4} \delta _{i_3 j_1} \delta _{i_2 j_3} \delta _{j_2 j_4}\\&+\delta _{i_1 i_3} \delta _{i_4 j_1} \delta _{i_2 j_3} \delta _{j_2 j_4}+\delta _{i_2 i_4}
   \delta _{i_1 j_1} \delta _{i_3 j_3} \delta _{j_2 j_4}+\delta _{i_1 i_4} \delta _{i_2 j_1} \delta _{i_3 j_3} \delta _{j_2 j_4}\\&+\delta _{i_2 i_3} \delta _{i_1
   j_1} \delta _{i_4 j_3} \delta _{j_2 j_4}+\delta _{i_1 i_3} \delta _{i_2 j_1} \delta _{i_4 j_3} \delta _{j_2 j_4}-\frac{128}{N^3} (\delta _{i_1 i_2} \delta _{i_3 i_4}
   \delta _{j_1 j_2} \delta _{j_3 j_4}\big)\\&-\frac{4}{N}\big( \delta _{i_2 i_4} \delta _{j_1 j_2} \delta _{i_3 j_3} \delta _{i_1 j_4}+\delta _{i_2 i_3} \delta _{j_1 j_2}
   \delta _{i_4 j_3} \delta _{i_1 j_4}+ \delta _{i_3 i_4} \delta _{i_2 j_2} \delta _{j_1 j_3} \delta _{i_1 j_4}+ \delta _{i_3 i_4} \delta _{i_2 j_1} \delta _{j_2
   j_3} \delta _{i_1 j_4}\\&+ \delta _{i_1 i_4} \delta _{j_1 j_2} \delta _{i_3 j_3} \delta _{i_2 j_4}+ \delta _{i_1 i_3} \delta _{j_1 j_2} \delta _{i_4 j_3} \delta
   _{i_2 j_4}+ \delta _{i_3 i_4} \delta _{i_1 j_2} \delta _{j_1 j_3} \delta _{i_2 j_4}+ \delta _{i_3 i_4} \delta _{i_1 j_1} \delta _{j_2 j_3} \delta _{i_2 j_4}\\&+
   \delta _{i_2 i_4} \delta _{j_1 j_2} \delta _{i_1 j_3} \delta _{i_3 j_4}+ \delta _{i_1 i_4} \delta _{j_1 j_2} \delta _{i_2 j_3} \delta _{i_3 j_4}+ \delta _{i_1
   i_2} \delta _{i_4 j_2} \delta _{j_1 j_3} \delta _{i_3 j_4}+ \delta _{i_1 i_2} \delta _{i_4 j_1} \delta _{j_2 j_3} \delta _{i_3 j_4}\\&+ \delta _{i_2 i_3} \delta
   _{j_1 j_2} \delta _{i_1 j_3} \delta _{i_4 j_4}+ \delta _{i_1 i_3} \delta _{j_1 j_2} \delta _{i_2 j_3} \delta _{i_4 j_4}+ \delta _{i_1 i_2} \delta _{i_3 j_2}
   \delta _{j_1 j_3} \delta _{i_4 j_4}+ \delta _{i_1 i_2} \delta _{i_3 j_1} \delta _{j_2 j_3} \delta _{i_4 j_4}\\&+ \delta _{i_3 i_4} \delta _{i_2 j_2} \delta _{i_1
   j_3} \delta _{j_1 j_4}+ \delta _{i_3 i_4} \delta _{i_1 j_2} \delta _{i_2 j_3} \delta _{j_1 j_4}+ \delta _{i_1 i_2} \delta _{i_4 j_2} \delta _{i_3 j_3} \delta
   _{j_1 j_4}+ \delta _{i_1 i_2} \delta _{i_3 j_2} \delta _{i_4 j_3} \delta _{j_1 j_4}\\&+2 \delta _{i_2 i_3} \delta _{i_1 i_4} \delta _{j_2 j_3} \delta _{j_1 j_4}+2
   \delta _{i_1 i_3} \delta _{i_2 i_4} \delta _{j_2 j_3} \delta _{j_1 j_4}+ \delta _{i_3 i_4} \delta _{i_2 j_1} \delta _{i_1 j_3} \delta _{j_2 j_4}+ \delta _{i_3
   i_4} \delta _{i_1 j_1} \delta _{i_2 j_3} \delta _{j_2 j_4}\\&+ \delta _{i_1 i_2} \delta _{i_4 j_1} \delta _{i_3 j_3} \delta _{j_2 j_4}+ \delta _{i_1 i_2} \delta
   _{i_3 j_1} \delta _{i_4 j_3} \delta _{j_2 j_4}+2 \delta _{i_2 i_3} \delta _{i_1 i_4} \delta _{j_1 j_3} \delta _{j_2 j_4}+2 \delta _{i_1 i_3} \delta _{i_2 i_4}
   \delta _{j_1 j_3} \delta _{j_2 j_4}\\&+ \delta _{i_2 i_4} \delta _{i_3 j_1} \delta _{i_1 j_2} \delta _{j_3 j_4}+ \delta _{i_2 i_3} \delta _{i_4 j_1} \delta _{i_1
   j_2} \delta _{j_3 j_4}+ \delta _{i_1 i_4} \delta _{i_3 j_1} \delta _{i_2 j_2} \delta _{j_3 j_4}+ \delta _{i_1 i_3} \delta _{i_4 j_1} \delta _{i_2 j_2} \delta
   _{j_3 j_4}\\&+ \delta _{i_2 i_4} \delta _{i_1 j_1} \delta _{i_3 j_2} \delta _{j_3 j_4}+ \delta _{i_1 i_4} \delta _{i_2 j_1} \delta _{i_3 j_2} \delta _{j_3 j_4}+
   \delta _{i_2 i_3} \delta _{i_1 j_1} \delta _{i_4 j_2} \delta _{j_3 j_4}+ \delta _{i_1 i_3} \delta _{i_2 j_1} \delta _{i_4 j_2} \delta _{j_3 j_4}\big)\\&+\frac{16}{N^2}\big(
   \delta _{i_3 i_4} \delta _{j_1 j_2} \delta _{i_2 j_3} \delta _{i_1 j_4}+ \delta _{i_3 i_4} \delta _{j_1 j_2} \delta _{i_1 j_3} \delta _{i_2 j_4}+ \delta
   _{i_1 i_2} \delta _{j_1 j_2} \delta _{i_4 j_3} \delta _{i_3 j_4}+ \delta _{i_1 i_2} \delta _{j_1 j_2} \delta _{i_3 j_3} \delta _{i_4 j_4}\\&+2 \delta _{i_1 i_2}
   \delta _{i_3 i_4} \delta _{j_2 j_3} \delta _{j_1 j_4}+2 \delta _{i_1 i_2} \delta _{i_3 i_4} \delta _{j_1 j_3} \delta _{j_2 j_4}+ \delta _{i_3 i_4} \delta
   _{i_2 j_1} \delta _{i_1 j_2} \delta _{j_3 j_4}+ \delta _{i_3 i_4} \delta _{i_1 j_1} \delta _{i_2 j_2} \delta _{j_3 j_4}\\&+ \delta _{i_1 i_2} \delta _{i_4 j_1}
   \delta _{i_3 j_2} \delta _{j_3 j_4}+ \delta _{i_1 i_2} \delta _{i_3 j_1} \delta _{i_4 j_2} \delta _{j_3 j_4}+2 \delta _{i_2 i_3} \delta _{i_1 i_4} \delta
   _{j_1 j_2} \delta _{j_3 j_4}+2 \delta _{i_1 i_3} \delta _{i_2 i_4} \delta _{j_1 j_2} \delta _{j_3 j_4}).
\end{align*}

Matching tensor structures between the $(12)-(34)$ channel and the $(14)-(32)$ channel gives the crossing relations:
\be
\left(\frac{u}{v}\right)^{\Delta_{\Phi}} G_{r,t}(v,u) = \cM_r^{r'} G_{r',s}(u,v),
\ee
with
\be
\cM^{r'}_{r} &=& \left(
\setlength\arraycolsep{1pt}\begin{array}{cccccc}
 \frac{2}{(N-1) (N+2)} & \frac{4 N}{(N-1) (N+2)} & \frac{4 (N-2) (N+4)}{(N-1) N (N+2)} & \frac{(N-3) N (N+1)}{(N-1)^2 (N+2)} & \frac{(N-2) (N+1) (N+4)}{(N-1)
   (N+2)^2} & \frac{N (N+1) (N+6)}{(N-1) (N+2)^2} \\
 \frac{1}{2 (N+2)} & \frac{1}{2} & \frac{(N-2) (N+4)}{2 N (N+2)} & \frac{(N-3) (N+1)}{4 (N-1) (N+2)} & 0 & -\frac{(N+1) (N+6)}{2 (N+2)^2} \\
 \frac{N}{2 (N-2) (N+4)} & \frac{N^2}{2 (N-2) (N+4)} & \frac{N^2+4 N-24}{2 (N-2) (N+4)} & -\frac{(N-3) N^2 (N+1)}{4 (N-2)^2 (N-1) (N+4)} & -\frac{N (N+1)}{(N-2)
   (N+2) (N+4)} & \frac{N^2 (N+1) (N+6)}{2 (N-2) (N+2) (N+4)^2} \\
 \frac{1}{3} & \frac{2}{3} & -\frac{2 (N+4)}{3 N} & \frac{N^2-2 N+3}{3 (N-2) (N-1)} & -\frac{N+4}{3 (N+2)} & \frac{N+6}{3 (N+2)} \\
 \frac{1}{2} & 0 & -\frac{4}{N} & -\frac{(N-3) N}{2 (N-2) (N-1)} & \frac{1}{2} & -\frac{N (N+6)}{2 (N+2) (N+4)} \\
 \frac{1}{6} & -\frac{2}{3} & \frac{2 (N-2)}{3 N} & \frac{N-3}{6 (N-1)} & -\frac{N-2}{6 (N+2)} & \frac{(N-2) N}{6 (N+2) (N+4)} \\
\end{array}
\right).\nn\\
\ee

\subsection{$SU(N)$ Adjoints}

Expanding the 4-point function of $SU(N)$ adjoints $\<\Phi \Phi \Phi \Phi \>$, we take the tensor structures in (\ref{eq:adj4pt}) to be:
\be
t^{\Phi,I}&=& \ \delta _{i_2}^{j_1} \delta _{i_3}^{j_4} \delta _{i_4}^{j_3} \delta _{i_1}^{j_2}-\frac{1}{N}\left(\delta _{i_2}^{j_2} \delta _{i_3}^{j_4} \delta _{i_4}^{j_3} \delta _{i_1}^{j_1}+\delta _{i_2}^{j_1} \delta _{i_3}^{j_3} \delta
   _{i_4}^{j_4} \delta _{i_1}^{j_2}\right)+\frac{1}{N^2} \delta _{i_2}^{j_2} \delta _{i_3}^{j_3} \delta _{i_4}^{j_4} \delta
   _{i_1}^{j_1}, \nn\\
t^{\Phi,Adj_a}&=&\delta _{i_2}^{j_4} \delta _{i_4}^{j_3} \delta _{i_3}^{j_1}\delta _{i_1}^{j_2}+\delta _{i_2}^{j_3} \delta _{i_4}^{j_1} \delta _{i_3}^{j_4}\delta _{i_1}^{j_2}+\delta
   _{i_2}^{j_1} \delta _{i_3}^{j_4} \delta _{i_4}^{j_2} \delta _{i_1}^{j_3}+\delta _{i_2}^{j_1} \delta _{i_3}^{j_2} \delta _{i_4}^{j_3} \delta
   _{i_1}^{j_4}-\frac{16}{N^3} \delta _{i_2}^{j_2} \delta _{i_3}^{j_3} \delta _{i_4}^{j_4} \delta _{i_1}^{j_1} \nn\\
   &&+\frac{4}{N^2} \left(\left(2 \delta _{i_3}^{j_4} \delta
   _{i_4}^{j_3} \delta _{i_2}^{j_2}+\delta _{i_3}^{j_2} \delta _{i_4}^{j_4} \delta _{i_2}^{j_3}+\delta _{i_3}^{j_3} \delta _{i_4}^{j_2} \delta _{i_2}^{j_4}\right)
   \delta _{i_1}^{j_1} \right.\nn\\
   &&\left.\qquad+2 \delta _{i_2}^{j_1} \delta _{i_3}^{j_3} \delta _{i_4}^{j_4} \delta _{i_1}^{j_2}+\delta _{i_2}^{j_2} \delta _{i_3}^{j_1} \delta
   _{i_4}^{j_4} \delta _{i_1}^{j_3}+\delta _{i_2}^{j_2} \delta _{i_3}^{j_3} \delta _{i_4}^{j_1} \delta _{i_1}^{j_4}\right) \nn\\
   &&-\frac{2}{N} \left(\left(\delta
   _{i_2}^{j_4} \delta _{i_4}^{j_3} \delta _{i_3}^{j_2}+\delta _{i_2}^{j_3} \delta _{i_4}^{j_2} \delta _{i_3}^{j_4}\right) \delta _{i_1}^{j_1}+\left(2 \delta
   _{i_3}^{j_4} \delta _{i_4}^{j_3} \delta _{i_2}^{j_1}+\delta _{i_3}^{j_1} \delta _{i_4}^{j_4} \delta _{i_2}^{j_3}+\delta _{i_3}^{j_3} \delta _{i_4}^{j_1} \delta
   _{i_2}^{j_4}\right) \delta _{i_1}^{j_2} \right.\nn\\
   &&\left.\qquad+\left(\delta _{i_3}^{j_2} \delta _{i_4}^{j_4} \delta _{i_2}^{j_1}+\delta _{i_3}^{j_4} \delta _{i_4}^{j_1} \delta
   _{i_2}^{j_2}\right) \delta _{i_1}^{j_3}+\left(\delta _{i_3}^{j_3} \delta _{i_4}^{j_2} \delta _{i_2}^{j_1}+\delta _{i_3}^{j_1} \delta _{i_4}^{j_3} \delta
   _{i_2}^{j_2}\right) \delta _{i_1}^{j_4}\right), \nn\\
t^{\Phi,Adj_s}&=&\delta _{i_1}^{j_4} \delta _{i_3}^{j_2} \delta _{i_4}^{j_3} \delta
   _{i_2}^{j_1}-\delta _{i_1}^{j_3} \delta _{i_3}^{j_4} \delta _{i_4}^{j_2} \delta _{i_2}^{j_1}+ \delta _{i_1}^{j_2}\delta _{i_2}^{j_3} \delta _{i_3}^{j_4} \delta _{i_4}^{j_1}-\delta _{i_1}^{j_2}\delta _{i_2}^{j_4} \delta _{i_3}^{j_1} \delta
   _{i_4}^{j_3},\nn\\
   t^{\Phi,(S,\bar{A})_a+(A,\bar{S})_a}&=&\delta _{i_1}^{j_4} \delta _{i_2}^{j_3} \delta _{i_4}^{j_1} \delta
   _{i_3}^{j_2}-\delta _{i_1}^{j_3} \delta _{i_2}^{j_4} \delta _{i_4}^{j_2} \delta _{i_3}^{j_1} \nn\\
   &&+\frac{1}{N} \left(\left(\delta _{i_2}^{j_4} \delta _{i_3}^{j_3} \delta _{i_4}^{j_2}-\delta _{i_2}^{j_3} \delta _{i_3}^{j_2} \delta _{i_4}^{j_4}\right) \delta 
   _{i_1}^{j_1} + \left(\delta _{i_2}^{j_4} \delta _{i_3}^{j_1} \delta _{i_4}^{j_3}-\delta _{i_2}^{j_3} \delta _{i_3}^{j_4} \delta _{i_4}^{j_1}\right) \delta
   _{i_1}^{j_2}\right.\nn\\ 
   &&\left.\qquad+\left(\delta _{i_3}^{j_4} \delta _{i_4}^{j_2} \delta _{i_2}^{j_1}+\delta _{i_3}^{j_1} \delta _{i_4}^{j_4} \delta _{i_2}^{j_2}\right) \delta
   _{i_1}^{j_3}-\left(\delta _{i_3}^{j_2} \delta _{i_4}^{j_3} \delta _{i_2}^{j_1}+\delta _{i_3}^{j_3} \delta _{i_4}^{j_1} \delta _{i_2}^{j_2}\right) \delta
   _{i_1}^{j_4}\right),\nn\\
   t^{\Phi,(A,\bar{A})_s} &=&\left(\delta _{i_1}^{j_4} \delta _{i_2}^{j_3}-\delta _{i_1}^{j_3} \delta _{i_2}^{j_4}\right) \left(\delta _{i_3}^{j_2} \delta _{i_4}^{j_1}-\delta _{i_3}^{j_1}
   \delta _{i_4}^{j_2}\right) \nn\\
   &&+\frac{1}{N-2}\left(\left(\left(\delta _{i_3}^{j_4} \delta _{i_4}^{j_2}-\delta _{i_3}^{j_2} \delta _{i_4}^{j_4}\right) \delta
   _{i_2}^{j_3}+\left(\delta _{i_3}^{j_2} \delta _{i_4}^{j_3}-\delta _{i_3}^{j_3} \delta _{i_4}^{j_2}\right) \delta _{i_2}^{j_4}\right) \delta
   _{i_1}^{j_1} \right.\nn\\
   &&\left.
   \qquad\qquad+\left(\left(\delta _{i_3}^{j_1} \delta _{i_4}^{j_4}-\delta _{i_3}^{j_4} \delta _{i_4}^{j_1}\right) \delta _{i_2}^{j_3}+\left(\delta _{i_3}^{j_3}
   \delta _{i_4}^{j_1}-\delta _{i_3}^{j_1} \delta _{i_4}^{j_3}\right) \delta _{i_2}^{j_4}\right) \delta _{i_1}^{j_2}
   \right.\nn\\
   &&\left. \qquad\qquad  +\left(\left(\delta _{i_3}^{j_2} \delta
   _{i_4}^{j_4}-\delta _{i_3}^{j_4} \delta _{i_4}^{j_2}\right) \delta _{i_2}^{j_1}+\left(\delta _{i_3}^{j_4} \delta _{i_4}^{j_1}-\delta _{i_3}^{j_1} \delta
   _{i_4}^{j_4}\right) \delta _{i_2}^{j_2}\right) \delta _{i_1}^{j_3} \right.\nn\\
 &&\left.\qquad\qquad  
   +\left(\left(\delta _{i_3}^{j_3} \delta _{i_4}^{j_2}-\delta _{i_3}^{j_2} \delta
   _{i_4}^{j_3}\right) \delta _{i_2}^{j_1}+\left(\delta _{i_3}^{j_1} \delta _{i_4}^{j_3}-\delta _{i_3}^{j_3} \delta _{i_4}^{j_1}\right) \delta _{i_2}^{j_2}\right)
   \delta _{i_1}^{j_4} \right.\nn\\
   &&\left.\qquad\qquad+\frac{2}{N-1} \left(\delta _{i_1}^{j_2} \delta _{i_2}^{j_1}-\delta _{i_1}^{j_1} \delta _{i_2}^{j_2}\right) \left(\delta _{i_3}^{j_4} \delta
   _{i_4}^{j_3}-\delta _{i_3}^{j_3} \delta _{i_4}^{j_4}\right)\right),\nn\\
 t^{\Phi,(S,\bar{S})_s}&=&\left(\delta _{i_2}^{j_4} \delta _{i_1}^{j_3}+\delta _{i_2}^{j_3} \delta _{i_1}^{j_4}\right) \left(\delta _{i_4}^{j_2} \delta _{i_3}^{j_1}+\delta _{i_4}^{j_1}
   \delta _{i_3}^{j_2}\right) \nn\\
   &&-\frac{1}{N+2} \left(\left(\left(\delta _{i_4}^{j_4} \delta _{i_3}^{j_2}+\delta _{i_4}^{j_2} \delta _{i_3}^{j_4}\right) \delta
   _{i_2}^{j_3}+\left(\delta _{i_4}^{j_3} \delta _{i_3}^{j_2}+\delta _{i_4}^{j_2} \delta _{i_3}^{j_3}\right) \delta _{i_2}^{j_4}\right) \delta
   _{i_1}^{j_1} \right.\nn\\
&&\left.\qquad\qquad   
   +\left(\left(\delta _{i_4}^{j_4} \delta _{i_3}^{j_1}+\delta _{i_4}^{j_1} \delta _{i_3}^{j_4}\right) \delta _{i_2}^{j_3}+\left(\delta _{i_4}^{j_3}
   \delta _{i_3}^{j_1}+\delta _{i_4}^{j_1} \delta _{i_3}^{j_3}\right) \delta _{i_2}^{j_4}\right) \delta _{i_1}^{j_2}\right.\nn\\
&&\left.\quad\qquad   
   +\left(\left(\delta _{i_4}^{j_4} \delta
   _{i_3}^{j_2}+\delta _{i_4}^{j_2} \delta _{i_3}^{j_4}\right) \delta _{i_2}^{j_1}+\left(\delta _{i_4}^{j_4} \delta _{i_3}^{j_1}+\delta _{i_4}^{j_1} \delta
   _{i_3}^{j_4}\right) \delta _{i_2}^{j_2}\right) \delta _{i_1}^{j_3} \right.\nn\\
&&\left.\qquad\qquad  
   +\left(\left(\delta _{i_4}^{j_3} \delta _{i_3}^{j_2}+\delta _{i_4}^{j_2} \delta
   _{i_3}^{j_3}\right) \delta _{i_2}^{j_1}+\left(\delta _{i_4}^{j_3} \delta _{i_3}^{j_1}+\delta _{i_4}^{j_1} \delta _{i_3}^{j_3}\right) \delta _{i_2}^{j_2}\right)
   \delta _{i_1}^{j_4} \right.\nn\\
  && \left.\qquad\qquad-\frac{2}{N+1} \left(\delta _{i_2}^{j_2} \delta _{i_1}^{j_1}+\delta _{i_2}^{j_1} \delta _{i_1}^{j_2}\right) \left(\delta _{i_4}^{j_4} \delta
   _{i_3}^{j_3}+\delta _{i_4}^{j_3} \delta _{i_3}^{j_4}\right)\right).
\ee

Matching the index structures between the $(12)-(34)$ channel and the $(14)-(32)$ channel, we obtain the crossing relations relating the t-channel functions to the s-channel functions, given by:
\be
\left(\frac{u}{v}\right)^{\Delta_{\Phi}} G_{r,t}(v,u) = \cM_r^{r'} G_{r',s}(u,v)
\ee
with
\be
\cM^{r'}_{r} &=&
\left(
\setlength\arraycolsep{1pt}\begin{array}{cccccc}
 \frac{1}{(N-1) (N+1)} & \frac{2 N}{(N-1) (N+1)} & \frac{2 (N-2) (N+2)}{(N-1) N (N+1)} & \frac{(N-2) (N+2)}{(N-1) (N+1)} & \frac{(N-3) N^2}{(N-1)^2 (N+1)} & \frac{N^2 (N+3)}{(N-1) (N+1)^2} \\
 \frac{1}{2 N} & \frac{1}{2} & \frac{(N-2) (N+2)}{2 N^2} & 0 & \frac{N-3}{2 (N-1)} & -\frac{N+3}{2 (N+1)} \\
 \frac{N}{2 (N-2) (N+2)} & \frac{N^2}{2 (N-2) (N+2)} & \frac{N^2-12}{2 (N-2) (N+2)} & -\frac{N}{(N-2) (N+2)} & -\frac{(N-3) N^3}{2 (N-2)^2 (N-1) (N+2)} & \frac{N^3 (N+3)}{2 (N-2) (N+1) (N+2)^2}
   \\
 \frac{1}{2} & 0 & -\frac{2}{N} & \frac{1}{2} & -\frac{(N-3) N}{2 (N-2) (N-1)} & -\frac{N (N+3)}{2 (N+1) (N+2)} \\
 \frac{1}{4} & \frac{1}{2} & -\frac{N+2}{2 N} & -\frac{N+2}{4 N} & \frac{N^2-N+2}{4 (N-2) (N-1)} & \frac{N+3}{4 (N+1)} \\
 \frac{1}{4} & -\frac{1}{2} & \frac{N-2}{2 N} & -\frac{N-2}{4 N} & \frac{N-3}{4 (N-1)} & \frac{N^2+N+2}{4 (N+1) (N+2)} \\
\end{array}
\right)\nn\\
\ee
in the basis 
\be
r = \bigg(I\,, Ads_a\,, Adj_s\,, (S,\bar{A})_a \oplus (A,\bar{S})_a\,, (A,\bar{A})_s\,, (S,\bar{S})_s\bigg).
\ee

When $N=3$ the $(A,\bar{A})_s$ representation does not exist and the equation reduces to
\be
\left(\frac{u}{v}\right)^{\Delta_{\Phi}} \left( \begin{array}{c} G_{I,t}(v,u) \\ G_{Adj_a,t}(v,u) \\ G_{Adj_s,t}(v,u) \\ G_{(S,\bar{A})_a,t}(v,u) \\ G_{(S,\bar{S})_s,t}(v,u) \end{array}\right) = \left(
\begin{array}{ccccc}
 \frac{1}{8} & \frac{3}{4} & \frac{5}{12} & \frac{5}{8}  & \frac{27}{16} \\
 \frac{1}{6} & \frac{1}{2} & \frac{5}{18} & 0  & -\frac{3}{4} \\
 \frac{3}{10} & \frac{9}{10} & -\frac{3}{10} & -\frac{3}{5}  & \frac{81}{100} \\
 \frac{1}{2} & 0 & -\frac{2}{3} & \frac{1}{2}  & -\frac{9}{20} \\
 \frac{1}{4} & -\frac{1}{2} & \frac{1}{6} & -\frac{1}{12}  & \frac{7}{40} \\
\end{array}
\right) \times \left( \begin{array}{c} G_{I,s}(u,v) \\ G_{Adj_a,s}(u,v) \\ G_{Adj_s,s}(u,v) \\ G_{(S,\bar{A})_a,s}(u,v) \\ G_{(S,\bar{S})_s,s}(u,v) \end{array}\right), 
\ee
while in the case that $N=2$ we only have the $I$, $Adj_a$, and $(S,\bar{S})_s$ representations with
\be
\left(\frac{u}{v}\right)^{\Delta_{\Phi}} \left( \begin{array}{c} G_{I,t}(v,u) \\ G_{Adj_a,t}(v,u) \\ G_{(S,\bar{S})_s,t}(v,u) \end{array}\right) = 
\left(
\begin{array}{cccccc}
 \frac{1}{3} & \frac{4}{3} & \frac{20}{9} \\
 \frac{1}{4} & \frac{1}{2} & -\frac{5}{6} \\
 \frac{1}{4} & -\frac{1}{2} & \frac{1}{6} \\
\end{array}
\right)
\times \left( \begin{array}{c} G_{I,s}(u,v) \\ G_{Adj_a,s}(u,v) \\ G_{(S,\bar{S})_s,s}(u,v) \end{array}\right).
\ee

\section{Crossing Symmetry of Leading $\log(v)$ Terms}
\label{app:LargeNArgument}

In this Appendix, we show that it is impossible to reproduce the $\log v$ term on the RHS of (\ref{eq:LargeNCrossing3}) in a unitary CFT. 
We can decompose $I_{t}(v,u)$ into contributions from operators with different twists. At leading order in $1/N,\ u,$ and $v$, the crossing equation becomes
\begin{align}
-\frac{1}{2C_{J}S_{d}^{2}}\frac{\Gamma(d)}{\Gamma(\frac{d}{2})^{2}}\log v & =v^{-\Delta_{\phi}}u^{\Delta_{\phi}-\frac{d-2}{2}}{\displaystyle \sum_{\tau,\ell}}P_{\tau,\ell}g_{\tau,\ell}(v,u),\nn\\
 & ={\displaystyle \sum_{\tau,\ell}}\left(\lim_{u\rightarrow0}u^{\Delta_{\phi}-\frac{d-2}{2}}P_{\tau,\ell}k_{2\ell}(1-u)\right)v^{\frac{\tau}{2}-\Delta_{\phi}}F^{(d)}(\tau,0),\nn\\
 & =\displaystyle{\int_{\frac{d-2}{2}}^{\infty}} d\sigma\rho(\sigma)v^{\frac{\sigma}{2}-\Delta_{\phi}}F^{(d)}(\sigma,0).\label{eq:LargeNCrossing2}
\end{align}
We used the form of the conformal block at small $u$ and $v$:
\bea
g_{\tau,\ell}(v,u)&\approx& k_{2\ell}(1-u)v^{\frac{\tau}{2}}F^{(d)}(\tau,v), \nn\\
k_{2\beta}(x)&\equiv& x^{\beta} \ _{2}F_{1}(\beta,\beta,2\beta,x),
\eea
and defined the density function
\begin{equation}
\rho(\sigma)\equiv\lim_{u\rightarrow0}u^{\Delta_{\phi}-\frac{d-2}{2}}{\displaystyle \sum_{\tau,\ell}}P_{\tau,\ell}k_{2\ell}(1-u)\delta(\tau-\sigma).
\end{equation}
Note that this decomposition is an integral over positive contributions, since $F^{(d)}(\sigma,v)$ is positive and analytic near $v\rightarrow0$ and $\rho(\sigma)\ge0$ in unitary CFTs for this 4-point function.

We first show that the tail of the integral in (\ref{eq:LargeNCrossing2}) cannot give rise the the $\log v$ term on the LHS. This part of the argument parallels the argument given in Appendix B.2 in \cite{Fitzpatrick:2012yx}. In particular, we study the integral over operators with twists higher than $\tau^* \gg 1$. Choosing a constant $1<\lambda<\frac{1}{v}$, we have
\begin{align}
\displaystyle{\int_{\tau^*}^{\infty}} d\sigma\rho(\sigma)v^{\frac{\sigma}{2}-\Delta_{\phi}}F^{(d)}(\sigma,0) &\le \lambda^{-\frac{\tau^*}{2}} \displaystyle{\int_{\tau^*}^{\infty}} d\sigma\rho(\sigma)(\lambda v)^{\frac{\sigma}{2}-\Delta_{\phi}}F^{(d)}(\sigma,0), \nn\\
&\le \lambda^{-\frac{\tau^*}{2}} \displaystyle{\int_{\frac{d-2}{2}}^{\infty}} d\sigma\rho(\sigma)(\lambda v)^{\frac{\sigma}{2}-\Delta_{\phi}}F^{(d)}(\sigma,0),\nn\\
& =\frac{1}{2C_{J}S_{d}^{2}}\frac{\Gamma(d)}{\Gamma(\frac{d}{2})^{2}}\lambda^{-\frac{\tau_{*}}{2}}\left(-\log\lambda v\right),
\end{align}
where we used the positivity of the integrand in the first two lines, and crossing symmetry at the point $(u,\lambda v)$ in the last line. The condition $\lambda<\frac{1}{v}$ follows from the convergence of the t-channel conformal block decomposition. 
Now we can choose $\lambda=\frac{1}{2v}$, obtaining
\begin{equation}
\displaystyle{\int_{\tau^*}^{\infty}} d\sigma\rho(\sigma)v^{\frac{\sigma}{2}-\Delta_{\phi}}F^{(d)}(\sigma,0)\le\frac{1}{2C_{J}S_{d}^{2}}\frac{\Gamma(d)}{\Gamma(\frac{d}{2})^{2}}\left(-\log\frac{1}{2}\right)(2v)^{\frac{\tau_{*}}{2}}.
\end{equation}
So given a large but finite $\tau^{*}$, for any $0<v\ll1$, the sum of all operators with twists higher than $\tau^{*}$ is bounded by $\sim v^{\frac{\tau^{*}}{2}}$ and cannot generate a $\log v$ term with finite coefficient. 
Therefore, the $\log v$ can only come from a finite part $[\frac{d-2}{2},\tau^*]$ of the integration region.

When $\sigma<2\Delta_{\phi}$, $\rho=0$, otherwise the RHS of (\ref{eq:LargeNCrossing2}) will have a power law divergence when $v\rightarrow0$. The only way to reproduce the $\log v$ on the LHS would be to have
\begin{equation}
\rho(\sigma)=A_0 \partial_\sigma\delta(\sigma-2\Delta_{\phi})+\dots,
\end{equation}
where
\be
A_{0} =\frac{1}{C_{J}S_{d}^{2}}\frac{\Gamma(d)}{\Gamma(\frac{d}{2})^{2}}\frac{1}{F^{(d)}(2\Delta_{\phi},0)}>0.
\ee
More generally, to solve the crossing equation (\ref{eq:LargeNCorssing1}) at the leading order in $u$ and all orders in $v$, we would need a sum of the form
\begin{equation}
\rho(\sigma)={\displaystyle \sum_{k}A_{k}\partial_{\sigma}\delta(\sigma-2\Delta_{\phi}-2n)+B_{k}\delta(\sigma-2\Delta_{\phi}-2n)},
\end{equation}
where the coefficients can be determined order by order. 

This solution, however, violates unitarity because $\partial\delta(\sigma-2\Delta_{\phi})$ is not a positive distribution. 
In particular, there exist smooth positive functions $h(\sigma)>0$ such that $\int d\sigma\rho(\sigma)h(\sigma)<0$.
This implies that there are no consistent unitary CFTs satisfying the assumptions described at the beginning of Section \ref{subsec:largeN}.\footnote{A similar situation arises in the bootstrap of $\<\f_{1}\f_{2}\f_{1}\f_{2}\>$ involving two distinct real scalars. The identity operator does not appear in the s-channel, which is then dominated by a conformal block containing a $\log(v)$ term at the leading order in $u$. The key difference is that $\rho$ is not positive definite as there is an extra factor of $(-1)^{\ell}$ in the OPE coefficients, which produces the relative sign. The anomalous dimensions of even and odd spin double-twist operators also contains a $(-1)^\ell$ such that the even and odd spin towers generate a density $\rho \propto \partial\delta(\sigma - 2\Delta_\phi)$ after summing over $\ell$.}

\bibliography{Biblio}{}
\bibliographystyle{utphys}

\end{document}